# Topological Braiding of Bloch Eigenmodes Protected by Non-Abelian Quaternion Invariants

Xiao-Ming Wang[1,2], Jiaying Xu[1], Xulong Wang[1], Zhen Li[1], Guancong Ma[1,3, †]

[1]Department of Physics, Hong Kong Baptist University, Kowloon Tong, Hong Kong, China

[2]Department of Physics and Astronomy, The University of Tennessee, Knoxville, Tennessee 37996, USA

[3]Shenzhen Institute for Research and Continuing Education, Hong Kong Baptist University, Shenzhen 518000, China

**Abstract**

Braiding has attracted significant attention in physics because of its important role in describing the fundamental exchange of particles. Infusing the braiding with topological protection will make it robust against imperfections and perturbations, but such topological braiding is believed to be possible only in interacting quantum systems, e.g., topological superconductors. Here, we propose and demonstrate a new strategy of topological braiding that emerges from non-Abelian topological insulators, a class of recently discovered multi-band topological phase. We unveil a mathematical connection between braiding and non-Abelian quaternion invariants, by which Bloch eigenmodes under parallel transport produce braid sequences protected by the non-Abelian band topology. The braiding is also associated with geometric phases quantized over half the Brillouin zone. This new type of non-Abelian topological braiding is experimentally realized in acoustic systems with periodic synthetic dimensions. The results show that the principle discovered here is a new strategy towards topological braiding and can be extended for other types of classical waves and non-interacting quantum systems.

† Email: phgcma@hkbu.edu.hk

**Introduction**

Many physical phenomena are sensitive to the order of execution. Braiding is one prominent example, which concerns the intertwining of multiple strands by making crossings. When three or more strands are braided, the outcome patterns are dependent on the order of the crossings, and the set of braid operations are non-Abelian in character[1]. Non-Abelian braiding has garnered considerable interest due to its fundamental importance in physics and pivotal role in potential application in quantum computing[2,3]. Recently, various research has revealed diverse non-Abelian phenomena in classical systems[4–11]. In particular, non-Abelian braiding can emerge in the dynamic evolution of classical acoustic or photonic modes[12,13]. Such braiding dynamics are underpinned by the non-Abelian Berry phase (NABP)[14], the multi-band generalization of the U(1) Berry phase, accumulated in the adiabatic evolution of multiple non-topological states. So these braiding dynamics require fine-tuning of system parameters and succumbs to external perturbations. Topological braiding that is robust against perturbations is, therefore, highly sought after for its great potential in fault-tolerant applications. Perhaps the most investigated example is the non-Abelian anyons emerging as the edge excitations of topological phases under long-range entanglement. When two or more such topologically protected non-Abelian anyons are exchanged, their world lines form non-Abelian braids[3,15–17]. However, single-particle topological systems such as the quantum Hall effect[18], Chern insulators[19], and symmetry-protected topological phase[20], are categorized by Abelian topological invariants (i.e., single numbers) under AZ classification[21], and they do not exhibit non-Abelian characteristics. Recently, a series of groundbreaking developments show that parity-time (*PT*) symmetric multi-band systems exhibit unconventional topological properties beyond the AZ classification: the topological invariants form non-Abelian groups under homotopy classification[22,23]. Non-Abelian topological phases, including gapped insulators and semimetals, have been realized in photonic and acoustic crystals[23–28]. Theories have also predicted their natural appearance in spin-polarized magnetic materials such as ZrTe, $Na_3N$, and $C_3N_4$ [29,30]. Studies also revealed that the exchange of multiple topologically charged nodal points in the momentum space can swap their charges following non-Abelian braiding rules[29,31,32].

In this work, we reveal that *PT*-symmetry-protected, fully gapped, three-band non-Abelian topological phases, which are topologically characterized by the quaternion group $Q_8 \coloneqq \{\pm i, \pm j, \pm k, \pm 1 | i^2 = j^2 = k^2 = -1, ij = k, ji = -k\}$ with group elements denoted as $q_8$, are fundamentally linked to non-Abelian braiding in elegant ways. By examining the parallel transport of all three Bloch eigenmodes across the Brillouin zone (BZ), we observe that the NABPs can always be decomposed into the multiplications of the generators of Artin's braid group $B_3$. Each quaternion invariant is robustly associated with a topologically protected minimum braid sequence of the Bloch eigenmodes, which has direct observable effects. The braiding is protected by non-Abelian band topology, whereas the braided modes are not topological edge modes. The braiding phenomena are experimentally observed in a three-state acoustic system with periodic synthetic dimensions. The principle is realizable in other types of classical or quantum systems and can be generalized to obtain arbitrary topologically protected braid sequences involving more modes. Our findings open a new route towards topological non-Abelian braiding in non-interacting systems.

**Taking the square root of quaternion invariants.** Our study is based on a generic *PT*-symmetric three-band gapped Hamiltonian

$$H(q) = R(q)\mathcal{E}R^{\mathrm{T}}(q), \qquad (1)$$

where $R(q) \in \mathrm{SO}(3)$ is a rotational matrix, $q$ is the Bloch wave vector, $\mathcal{E} = \mathrm{diag}(f_1, f_2, f_3)$. Owing to the *PT*-symmetry, the eigenvectors of Eq. (1) are purely real, and they have a $\mathbb{Z}_2$ gauge freedom. The topological space of Eq. (1) is $X = \frac{O(3)}{\mathbb{Z}_2 \times \mathbb{Z}_2 \times \mathbb{Z}_2}$, and its fundamental homotopy group is $\pi_1(X) = Q_8$, i.e., the quaternion group. These are some of the main results in refs.[22,23], and are the starting point of our work.

The band topology of Eq. (1) can also be characterized using the NABP

$$W = \exp\left[\oint A_{3\times 3}(q)dq\right], \qquad (2)$$

where $A_{3\times 3}(q)$ is the non-Abelian Berry connection, i.e., a $3 \times 3$ unitary matrix with $a_{mn} = \langle \psi_m(q)|\partial_q|\psi_n(q)\rangle$ as entries, and $|\psi_{m,n}(q)\rangle$ denotes the Bloch wavefunctions with the subscripts indexing the bands. Note that owing to the *PT* symmetry, $A(q)$ is a real orthogonal matrix. There is direct correspondence between $W$ and the quaternion invariants, as summarized in Table 1, wherein $L_{x,y,z}$ are the three generators of $\mathrm{SO}(3)$. The results highlight the underlying connection between the $\mathrm{SO}(3)$ eigenvectors and the quaternion invariants. The *PT*-symmetry encoded in Eq. (1) further imposes that the rotation angle can only take quantized values of $0, \pm\pi, \pm 2\pi$ when $q$ traverses the entire BZ. The quaternion group, which is a subgroup of $\mathrm{SU}(2)$ and is widely used to describe 3D rotation, elegantly captures such characteristics, as depicted in Fig. 1(a). Because the rotations are due to the adiabatic evolutions across the BZ, NABP naturally emerges as the connection between the start and end states. These two pictures of eigenvector rotation and adiabatic evolution conform with and complement one another.

However, careful examination of the mapping between quaternion invariants and NABPs reveals two intriguing matters. First, there is a two-to-one mapping for all cases, i.e., invariants with opposite signs map to the same NABP. In other words, the NABP does not distinguish the differences between positive and negative invariants. An immediate question is whether the signs of the invariants are consequential. Second, rotation about any axis by $\pm 2\pi$ is mapped to the invariants $-1$. Naturally, we ask whether they are additional "hidden texture" for each quaternion invariant. (The invariant of $+1$ is topologically trivial and corresponds to no rotation, so the choice of rotation axis is irrelevant.) Our investigation not only provides answers to these questions but also unveils a route for topologically protected non-Abelian braiding.

The treatment starts by "taking the square root" of NABPs, i.e., splits the NABPs into two equal parts. Take the cases $Q_8 = \pm i$ as an example, which corresponds to NABPs $W = \exp(\pm \pi L_x)$. The square roots are $\exp\left(\pm \frac{\pi}{2} L_x\right)$. At first sight, this step seems totally trivial because it simply shows that the rotation of $\pi$ is equivalent to two sequential rotations by $\pi/2$ about the same axis, as shown in Fig. 1(e). However, an intriguing observation can be made

$$\exp\left(\pm \frac{\pi}{2} L_x\right) = \begin{pmatrix} 1 & 0 & 0 \\ 0 & 0 & \mp 1 \\ 0 & \pm 1 & 0 \end{pmatrix} \to b_{12}^{\pm 1}, \qquad (4)$$

where $b_{12}^{\pm 1}$ are two conjugate generators of the Artin's braiding group $B_3$. In other words, the $\pi/2$ rotation maps one-to-one to a braiding operation. This relation is depicted in Fig. 1(b). Likewise, it is straightforward to obtain $\exp\left(\pm \frac{\pi}{2} L_z\right) \to b_{23}^{\pm 1}$ for $Q_8 = \pm k$.

The mapping from $\pm j$ to braiding is slightly more complicate: $\exp\left(\pm \frac{\pi}{2} L_y\right) \to (b_{12} b_{23} b_{12}^{-1})^{\pm 1}$, which is ostensibly bizarre. However, this is a natural requirement enforced by the Lie algebra $\mathfrak{so}(3)$, $[L_z, L_x] = L_y$, which implies that $L_y$ must relate to both $L_x$ and $L_z$. Moreover, the equality $e^{\pi L_x/2} e^{\theta L_z} e^{-\pi L_x/2} = e^{\theta L_y}$ also indicates that, if the $\frac{\pi}{2}$ rotations about $x$ and $z$ axes only relate a single braid generator, then the same rotation about $y$ axis must be represented by a sequence of both generators. Also, the braid sequences for $\pm i, \pm k$ are commutative, so $\pm j$ must be a non-commutative sequence with different generators to underpin the non-Abelian characteristics. Indeed, the cases of $\pm i, \pm k$ alone can be fully described by degrading to SO(2) rotation in specific conditions, and $\pi_1(SO(2)) = \mathbb{Z}$ is Abelian.

Next, by the relation $i^2 = j^2 = k^2 = -1$, the invariant $-1$ corresponds to rotation about any axis by $\pm 2\pi$, from which we have $\exp(\pm 2\pi L_{x,y,z}) = \mathbf{1}$, with $\mathbf{1}$ being a $3 \times 3$ identity matrix. The braid sequences of rotation about $x, y, z$ axes are $(b_{12}^{\pm 1})^4$ for $i^2$, $(b_{12} b_{23} b_{12}^{-1})^4$ or $(b_{12}^{-1} b_{23}^{-1} b_{12})^4$ for $j^2$, and $(b_{23}^{\pm 1})^4$ for $k^2$. However, these cases are not exhaustive because rotation about any axis by $2\pi$ gives rise to the identity. In fact, it has been mathematically shown that all 3D rotations by $2\pi$ and $\pi$ can be decomposed into 48 different braid sequences in total[33]. Among these, several special cases are particularly interesting, e.g., $\exp(\pm 2\pi M_a) = (b_{23} b_{12} b_{23})^{\pm 2} = (b_{12} b_{23} b_{12})^{\pm 2}$, $\exp(\pm 2\pi M_b) = (b_{12}^{-1} b_{23} b_{12}^{-1})^{\pm 2}$, with

$$M_a = \frac{1}{\sqrt{2}} \begin{pmatrix} 0 & -1 & 0 \\ 1 & 0 & -1 \\ 0 & 1 & 0 \end{pmatrix}, M_b = \frac{1}{\sqrt{2}} \begin{pmatrix} 0 & 1 & 0 \\ -1 & 0 & -1 \\ 0 & 1 & 0 \end{pmatrix}, \qquad (5)$$

which are direct manifestations of the Yang-Baxter equation[34]. Following the above derivation, the relations among $Q_8$, $W$, and braid sequences are summarized in Table 1.

To summarize the discussion so far, we have rigorously shown that non-Abelian characteristics of the band topology and the NABPs, which are all diagonal matrices that always commute and do not display any non-Abelian characteristics. To see through this paradox and reveal the non-Abelian characteristics, it was necessary to lift SO(3) to SU(2), then mapping SU(2) to $Q_8$ as $\pm \sigma_0 \to \pm 1, \pm i\sigma_x \to \pm i, \pm i\sigma_y \to \pm j, \pm i\sigma_z \to \pm k$[22].

**Table 1** The relations between the quaternion invariants, NABPs, and braid sequences. The matrix representations of all cases of $-1$ are the $3 \times 3$ identity matrix. The quaternion invariants correspond to quantized rotation angles of $\pm \pi, \pm 2\pi$, so the braid sequences must have an even number of crossings. The braid sequence for the case with $\exp(\pm 2\pi L_{x,y,z}) = \mathbf{1}$ can be readily derived by repeating the braiding operation for $q_8 = \pm i, \pm j, \pm k$ twice. These cases are omitted for brevity. Only the irreducible braid sequences are listed.

| $q_8$ | Non-Abelian Berry phase $W$ | Braid word | Braid sequence |
|---|---|---|---|
| $i$ | $\exp(\pi L_x) = \text{diag}(1, -1, -1)$ | $b_{12} b_{12}$ | 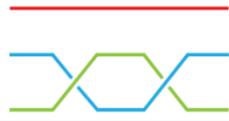 |
| $-i$ | $\exp(-\pi L_x) = \text{diag}(1, -1, -1)$ | $b_{12}^{-1} b_{12}^{-1}$ | 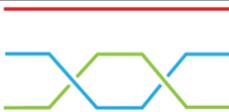 |
| $k$ | $\exp(\pi L_z) = \text{diag}(-1,-1,1)$ | $b_{23}b_{23}$ | 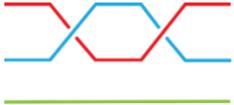 |
| $-k$ | $\exp(-\pi L_z) = \text{diag}(-1,-1,1)$ | $b_{23}^{-1}b_{23}^{-1}$ | 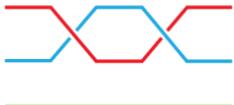 |
| $j$ | $\exp(\pi L_y) = \text{diag}(-1,1,-1)$ | $b_{12}b_{23}b_{12}^{-1}b_{12}b_{23}b_{12}^{-1}$ | 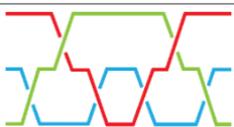 |
| $-j$ | $\exp(-\pi L_y) = \text{diag}(-1,1,-1)$ | $b_{12}b_{23}^{-1}b_{12}^{-1}b_{12}b_{23}^{-1}b_{12}^{-1}$ | 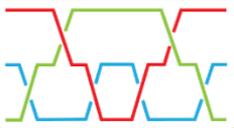 |
| | $\exp(\pm 2\pi L_{x,y,z}) = \mathbf{1}$ | — | — |
| $-\mathbf{1}$ | $\exp(2\pi M_a) = \mathbf{1}$ | $b_{23}b_{12}b_{23}b_{23}b_{12}b_{23}$ | 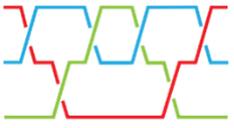 |
| | $\exp(2\pi M_b) = \mathbf{1}$ | $b_{12}^{-1}b_{23}b_{12}^{-1}b_{12}^{-1}b_{23}b_{12}^{-1}$ | 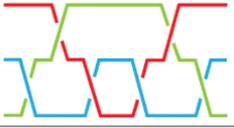 |
| | $\exp(2\pi M_a) = \mathbf{1}$ | $b_{12}b_{23}b_{12}b_{12}b_{23}b_{12}$ | 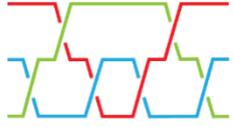 |

**Topological braiding of Bloch modes.** While the above results seem purely mathematical, they have observable consequences: the braiding of Bloch modes. This expands the observable effects of non-Abelian band topology beyond the quotient-type bulk-edge correspondence[23]. To demonstrate, we consider a concrete one-dimensional (1D) three-band Bloch Hamiltonian

$$H(q) = \begin{pmatrix} \omega_3 + 2v_3 \cos q & 2w_1 \sin q + 2w_2 \cos q & 0 \\ 2w_1 \sin q + 2w_2 \cos q & \omega_2 + 2v_2 \cos q & 2u \sin q \\ 0 & 2u \sin q & \omega_1 + 2v_1 \cos q \end{pmatrix}. \quad (6)$$

All parameters are real and are chosen such that the system is always gapped. The model respects $PT$-symmetry, so the three-band topology is captured by $Q_8$. It also has a hidden mirror symmetry

$$RH(q)R^{-1} = H(-q), \text{ with}$$

$$R = \begin{cases} \text{diag}(1,-1,1) & \text{for } w_1 \neq 0, w_2 = 0, \\ \text{diag}(1,1,-1) & \text{for } w_1 = 0, w_2 \neq 0. \end{cases} \quad (7)$$

The mirror symmetry ensures that the NABPs accumulated over each half of the BZ are the same, i.e., it guarantees the $\pm\pi/2$ rotation for $\pm i, \pm j, \pm k$ -phases ($\pm\pi$ for $-1$ -phases) completes exactly in half BZ. We set parameters $v_1 = 1, v_2 = 0, v_3 = -1, \omega_1 = \omega_2 = 0, \omega_3 = -4, w_1 = -1, w_2 = 0, u = 1$, such that the band topology is characterized by the invariant $Q_8 = k$, which is denoted as the $k$-phase for brevity. We trace the parallel transport of the three Bloch eigenmodes, denoted $|\psi_{1,2,3}(q)\rangle$, across the BZ. In the results shown in Fig. 2(a), the profiles of $|\psi_{2,3}(-\pi)\rangle$ are swapped with $|\psi_{2,3}(0)\rangle$, and $|\psi_3\rangle$ accumulates an additional phase factor of $\pi$, i.e., $\langle\psi_2(-\pi)|\psi_3(0)\rangle = -1$ and $\langle\psi_3(-\pi)|\psi_2(0)\rangle = 1$. $|\psi_1\rangle$ is identical at $q = -\pi$ and 0. The exchange of eigenmodes' profiles is clearly described by $b_{23}$. Notably, the braiding occurs without closing any gap or requiring modal degeneracy. Continuing the evolution from $q = 0$ to $\pi$, $|\psi_{2,3}\rangle$ swap their profiles again and $|\psi_2\rangle$ gains a $\pi$ phase, $|\psi_1\rangle$ remains unchanged. Upon completing the cyclic evolution, all three Bloch eigenmodes are recovered, but both $|\psi_{2,3}\rangle$ pick up an additional phase of $\pi$, which can also be attributed to a geometric phase.

**Experimental observation of the Bloch-modes braiding.** The model [Eq. (6)] can, in principle, be realized in non-interacting quantum systems, or classical systems such as photonics or metamaterials. Our goal is to precisely trace the Bloch eigenmodes, which is not straightforward in a finite-size specimen. Also, because the system is gapped, dynamic evolution will inevitably induce different dynamic phases for the three modes that have different eigenenergies, which makes the precise observation of the geometric phases difficult. To circumvent these complications, we choose to perform proof-of-principle experiments using a stroboscopic approach in a set of coupled acoustic cavities with a periodic synthetic dimension. The experimental setup is illustrated in Fig. 3(a). Simply put, the system consists of three annular acoustic cavities coupled by small waveguides. A thin metal plate is inserted in each cavity to reflect the circulating traveling waves, such that azimuthal standing-wave modes are formed. We focus on the first-order mode with an acoustic pressure profile $P(\theta) \propto \cos\theta$, where $\theta \in [0, 2\pi)$ naturally maps to a 1D BZ, so it is utilized as the synthetic dimension, i.e., $q \rightarrow \theta$. (For convenience, we choose $\theta \in [0, 2\pi]$ as the BZ in our experiments.) Examining Eq. (6), it is clear that $\theta$ needs to control the onsite resonant frequencies and nearest-neighboring coupling terms. The resonant frequency of each cavity is tunable by changing the azimuthal position of a metal block. The coupling strength among the cavities can be tuned by varying the azimuthal position of the connecting waveguides. Also, the coupling signs need to change. This is realized by connecting the cavities with additional longer waveguides. A detailed description of the acoustic setup is presented in the Methods section.

Figure 3(b) presents the simulated evolution of the pressure profiles of the eigenfunctions for the $k$-phase at selected values of $\theta$. System parameters are listed in the Supplementary Information. The unit of all parameters in the experimental system is rad·s$^{-1}$. To understand the results, we first examine the results at $\theta = 0$. For the second mode with the resonant frequency of $f_2$, the acoustic pressure is maximum (vanishing) at the middle (top and bottom) cavity, which directly corresponds to the eigenvector of $|\psi_2(0)\rangle = (0,1,0)^T$. Varying to $\theta = \frac{\pi}{2}$, all three cavities have non-zero acoustic pressure, signifying the evolution of the eigenvectors. At $\theta = \pi$, only the top cavity has large acoustic amplitude, which corresponds to $|\psi_2(\pi)\rangle = (1,0,0)^T$. As such, we can also see that the third mode at $f_3$ follows $|\psi_3(0)\rangle = (1,0,0)^T$ and $|\psi_3(\pi)\rangle = (0,-1,0)^T$, where the minus sign manifests as the additional $\pi$ phase in the middle cavity. The first mode at $f_1$ follows $|\psi_1(0)\rangle = (0,0,1)^T$ and $|\psi_1(\pi)\rangle = (0,0,1)^T$. In summary, the evolution of the three modes across half BZ is indeed described by $b_{23}$. In the second half BZ where $\theta$ varies from $\pi$ to $2\pi$, the acoustic eigenfunctions undergo the same evolution, resulting in $W = \exp(\pi L_z) = \text{diag}(-1,-1,1)$, which is equivalent to $b_{23}b_{23}$. The braiding of acoustic eigenfunctions is experimentally observed. The results are presented in Fig. 4(a-c).

The braiding pattern for the $j$-phase is less straightforward because all three modes are involved in the braiding process, as shown in Table 1. Under the parameters of the $j$-phase, the three eigenmodes are no longer localized in a single cavity at $\theta = 0$, as shown in Fig. 3(c). But the proper braiding can still be observed. The first eigenmode dwells in cavities 1 and 2, the third eigenmode is localized at cavity 3. At $\theta = \pi$, these two eigenmodes have exchanged their profiles. Meanwhile, the parity of the first eigenmode at $\theta = 0$ is changed compared with the third eigenmode at $\theta = \pi$. Because braiding operations only permit exchanging the nearest neighbors, the exchange of the first and third modes must be mediated by the second mode. Indeed, for the second eigenmodes, we observe that only cavities 1 and 2 have support at $\theta = 0$, but this distribution is not maintained in the intermediate steps, indicating its involvement in the braiding. This observation also conforms to the braid word for $j$ in Table 1.

Next, we examine the $-1$-phases. As mentioned, multiple choices of rotation axes and braid sequences are permitted for $q_8 = -1$ (Table 1). Here, we consider an interesting case with the evolution over half BZ ($\theta = 0$ to $\pi$) giving rise to the Yang-Baxter equation $b_{12}b_{23}b_{12} = b_{23}b_{12}b_{23}$. As shown in Fig. 3(d), the first and third modes exchange their profiles with the phases unchanged. But the second mode picks up a $\pi$ phase factor. The net outcome exactly maps to $b_{12}b_{23}b_{12}$ or $b_{23}b_{12}b_{23}$. This result is also verified experimentally, as shown by the data in Fig. 4(d-f).

**Quantized half-BZ geometric phases.** Observe that the braid operation not only swaps two Bloch modes' profiles but also flips the sign. Such sign changes are protected by a nontrivial topological invariant defined over half BZ using a relative homotopy group[35]. Simply put, owing to the mirror symmetry at $q = 0$ and $\pi$, the relative homotopy groups for our system [Eq. (6)] is $\pi_1(X, A) = \mathbb{Z}_2$, where $X$ is given after Eq. (1), $A \in X$ is topological space of $H(q)$ at the two mirror-symmetric points $q = \pm\pi, 0$. Note that mirror symmetry points exactly segment the BZ into two equal halves. So $\mathbb{Z}_2$ geometric phases can be defined over half BZ

$$\Theta_n^- = \text{Im}\log\langle\psi_n(-\pi)|\psi_m(0)\rangle, \Theta_n^+ = \text{Im}\log\langle\psi_m(0)|\psi_n(\pi)\rangle. \tag{8}$$

As indicated by the band indices $m$ and $n$, $\Theta_n^\pm$ are determined by comparing the identical eigenfunctions at the symmetry points, which can belong to different bands owing to the braiding discussed above. So $\Theta_n^\pm$ are fundamentally different from the standard the single-band Zak phase[36], which considers the full-BZ holonomy of a single band. Similar half-BZ quantization is also found in mirror-related Weyl points[37] and delicate topological insulators[38,39]. More detailed discussions are included in the Supplementary Information.

The results are indicated in Fig. 2. Careful examination reveals that $\Theta_n^\pm$ not only accounts for the sign flips during the braiding but also offer new insights into the non-Abelian band topology. To see this, notice that full-BZ Zak phases for both $\pm k$-phases are $\Theta = \Theta^- + \Theta^+ = \{\pi, \pi, 0\}$ for all bands [Fig. 2(a, d)]. But the half-BZ geometric phases for $k$-phase are $\Theta^- = \{0, \pi, 0\}$ and $\Theta^+ = \{0, 0, \pi\}$, whereas for $-k$, they are $\Theta^- = \{0, 0, \pi\}$ and $\Theta^+ = \{0, \pi, 0\}$. In other words, the sign of $Q_8$ corresponds to which half of the BZ are the quantized geometric phases accumulated. The similar distinction can be observed for other cases. In particular, $\Theta^\pm$ also clearly show that $q_8 = -1$ is topologically nontrivial, e.g., in Fig. 2(g), $\Theta_2^- = \Theta_2^+ = \pi$, which are clearly nontrivial. But the full-BZ Zak phase is $2\pi = 0$ under $\text{mod}(2\pi)$. All other scenarios of $q_8 = -1$ possess different combination of $\Theta^\pm$. The non-zero matrix elements in NABPs at $q = \pm\pi, 0$ are due to the non-trivial relative homotopy group, independent of the rotation axis and directions. This establishes a rigorous foundation for decomposing NABPs into sequences of braiding operations.

**Discussion.** Our work demonstrates a new paradigm for topologically protected braiding that hinges on the mathematical connection among NABPs, the eigenvector frame rotation, and braiding. The Bloch mode braiding is a fundamental feature of the non-Abelian topological insulator, so it is robust against local perturbations. This approach does not rely on the delicate edge excitations in interacting quantum systems. It can be dynamically implemented using the Floquet approach that maps Bloch momentum to periodic time modulation[40,41,32] or using the non-adiabatic holonomic approach[42,43]. The principle is realizable in other types of classical-wave systems, photonics, magnons, and single-particle quantum systems such as nitrogen vacancies in diamond.

From the results shown in Table 1, it appears that the possibilities of braid sequences are limited to a few choices. This limitation is due to the strong constraint imposed by the mirror symmetry. Relaxing the mirror symmetry, the array of allowed braid sequences is vastly increased. (In this case, half-BZ geometric phases are no longer quantized.) The only requirement here is that the total braiding operations are an even number, and the extra braid operations cancel each other over the full BZ. More generally, we consider a generic braid sequence $b_{12}^{\chi_1} b_{23}^{\chi_2} \ldots b_{12}^{\chi_{n-1}} b_{23}^{\chi_n}$, where $\chi_i$ is an integer. If the braiding degree of each operator is an even number, the braid sequence must map to one of the quaternion invariants (Detailed proof is in the Supplementary Information). To produce braiding of more modes, the systems need to possess more bands[24]. These generalizations are further discussed in the Supplementary Information.

The braiding perspective also refines the non-Abelian topological categories. Again, take the $k$-phase as an example. It is separated into two equal braid sequences, one for each half-BZ. This result is due to the mirror symmetry. Yet the quantized quaternion invariants exist even when mirror symmetry is broken – they are only underpinned by the $PT$ symmetry. So by relaxing the mirror symmetry, it is also possible to obtain different sequences from the $k$-phase like $b_{12}^4 b_{23}^{-2}$. Such distinction in braid sequences may be classified as new non-Abelian topological sub-genres. This will be a focus of future investigations.

**Methods**

**NABP and the mapping to the quaternion group.** The NABP is a multi-band generalization of the Berry phase. For a three-band system, it is defined as

$$W = \exp[\oint A_{3\times 3}(q) dq], \tag{A1}$$

where $A_{3\times 3}$ is the matrix-value Berry connection with $A_{mn} = \langle \psi_m(q) | \partial_q | \psi_n(q) \rangle$ as entries, where

$|\psi_n(q)\rangle$ denotes the Bloch eigenvector of the *n*-th band. We first decompose the Berry connection into a linear combination of SO(3) generators with

$$A_{3\times3}(q) = \sum_\mu \beta_\mu(q) L_\mu, \tag{A2}$$

where the summation runs through $\mu = x, y, z, 0$, with

$$L_x = \begin{pmatrix} 0 & 0 & 0 \\ 0 & 0 & -1 \\ 0 & 1 & 0 \end{pmatrix}, L_y = \begin{pmatrix} 0 & 0 & -1 \\ 0 & 0 & 0 \\ 1 & 0 & 0 \end{pmatrix}, L_z = \begin{pmatrix} 0 & -1 & 0 \\ 1 & 0 & 0 \\ 0 & 0 & 0 \end{pmatrix}, \tag{A3}$$

being the three generators of SO(3) and $L_0$ is a $3 \times 3$ identity. Then, lift the SO(3) to SU(2) by $L_\mu \to -\frac{i}{2}\sigma_\mu$, where $\sigma_\mu$ denote the Pauli matrices of spin-1/2.

$$A_{2\times2}(q) = -\frac{i}{2}\sum_\mu \beta_\mu(q)\sigma_\mu. \tag{A4}$$

And the NABP becomes

$$W_{2\times2} = \exp[\oint A_{2\times2}(q)dq]. \tag{A5}$$

After integration, $W_{2\times2}$ is non-commutative. The mapping between quaternion group and SU(2) group is given as $\pm\sigma_0 \to \pm 1, \pm i\sigma_x \to \pm i, \pm i\sigma_y \to \pm j, \pm i\sigma_z \to \pm k$.

**Acoustic experimental system and measurement.** The acoustic system consists of three coupled annular cavities filled with air. The cavities are stacked with a separation of 62 mm in between. The cavities have inner and outer radii of 8 mm and 35 mm, respectively, and a height of 28 mm. A metallic plate is placed in each cavity to reflect the clockwise and counterclockwise travelling waves such that azimuthal standing wave is formed. The resonant frequency of the first-order azimuthal mode is about 1400 Hz, with an acoustic pressure eigenfunction following $P(\theta) \propto \cos\theta$. Due to the sound-hard boundary condition (Neumann type), the anti-nodes of the acoustic pressure appear at the metallic plate, whose azimuthal position is defined as $\theta = 0$ $(2\pi)$. One wave node exists at $\theta = \pi$. In Eq. (6), the resonant frequency of each site (cavity) is tuned according to by $\sin\theta = \cos\left(\frac{\pi}{2} - \theta\right)$. This is achieved by placing a solid block in each cavity and varying their azimuthal positions together. The hopping terms in Eq. (6) follow functional relations of $\sin\theta$ and $\cos\theta$. They are realized by varying the azimuthal positions of the coupling tubes. Notably, the signs of some hopping terms change as $\theta$ varies from 0 to $2\pi$. So a set of two coupling tubes of different lengths, a short one at 62 mm and a long one at 124 mm, are used. According refs.[44,45], the short (long) tube realizes negative (positive) hopping. As such, the hopping terms can be continuously tuned across 0, which is confirmed in numerical simulations (see Supplementary Information for the results).

**Retrieval of the system parameters and the eigenvectors from the acoustic measurements.** The experiments we performed are the steady-state response measurements that yielded the response spectra at all three cavities. From the responses, the system parameters and the eigenvectors were obtained using the Green's function

$$G(\omega, \theta) = \sum_{n=1}^{3} \frac{G_0 |\psi_n(\theta)\rangle \langle \psi_n(\theta)|}{\omega - \omega_n(\theta) + i\gamma}, \tag{A6}$$

where $\omega_n$ is the resonant angular frequency of the $n$-th band, $\gamma$ represents the loss in the system, $G_0$ is a fitting constant. The steady-state response at cavity-$l$ ($l = 1,2,3$) is

$$P_l(\omega) = \langle p_l | G(\omega) | s \rangle, \tag{A7}$$

where $|s\rangle$ and $|p_l\rangle$ are $3 \times 1$ column vectors representing the source and probes. In our experiments, a loudspeaker is placed at the second cavity as the source, so $|s\rangle = (0,1,0)^\mathrm{T}$. A microphone is placed at each cavity as the probe, so $|p_1\rangle = (1,0,0)^\mathrm{T}$, $|p_2\rangle = (0,1,0)^\mathrm{T}$, $|p_1\rangle = (0,0,1)^\mathrm{T}$. To retrieve the system parameters (including $w_1, w_2, v_1, v_2, u, \gamma, \omega_n, A$) and the eigenvectors $|\psi_n(\theta)\rangle = (\varphi_1, \varphi_2, \varphi_3)^\mathrm{T}$, a least-square fitting is applied to fit the amplitudes and phases of the measured spectra $P_{1,2,3}(\omega)$.

**Acknowledgements.** X.-M. W. and G. M. thank Ruo-Yang Zhang for helpful discussion. X.-M. W. also acknowledges insightful discussion with Rui-Xing Zhang, Penghao Zhu and Jiabin Yu. This work was supported by the National Key R&D Program (2022YFA1404403), the Hong Kong Research Grants Council (RFS2223-2S01, 12301822), the National Natural Science Foundation of China (12472088), and the Hong Kong Baptist University (RC-RSRG/23-24/SCI/01 and RC-SFCRG/23-24/R2/SCI/12). X.-M. W. acknowledges the support of the Graduate Teaching Assistant Scholarship from The University of Tennessee, Knoxville.

**Author contributions.** X.-M. W. developed the theory and performed calculations. X.-M. W., X. W., and Z. L. developed the experimental system. J. X. and X.-M. W. performed the acoustic measurement. X.-M. W. and G. M. analyzed the data and wrote the manuscript. G. M. initiated and supervised the research.

**Competing interests.** The authors declare no competing interests.

# References

1. Kassel, C. & Turaev, V. *Braid Groups*. vol. 247 (Springer Science & Business Media, 2008).

2. Kitaev, A. Yu. Fault-tolerant quantum computation by anyons. *Annals of Physics* **303**, 2–30 (2003).

3. Nayak, C., Simon, S. H., Stern, A., Freedman, M. & Das Sarma, S. Non-Abelian anyons and topological quantum computation. *Rev. Mod. Phys.* **80**, 1083–1159 (2008).

4. Yang, Y. *et al.* Non-Abelian physics in light and sound. *Science* **383**, eadf9621 (2024).

5. Yan, Q. *et al.* Non-Abelian gauge field in optics. *Adv. Opt. Photon.* **15**, 907 (2023).

6. Chen, Y. *et al.* Non-Abelian gauge field optics. *Nat Commun* **10**, 3125 (2019).

7. Noh, J. *et al.* Braiding photonic topological zero modes. *Nat. Phys.* **16**, 989–993 (2020).

8. Sun, Y.-K. *et al.* Non-Abelian Thouless pumping in photonic waveguides. *Nat. Phys.* **18**, 1080–1085 (2022).

9. You, O. *et al.* Observation of Non-Abelian Thouless Pump. *Physical Review Letters* **128**, 244302 (2022).

10. Wu, J. *et al.* Non-Abelian gauge fields in circuit systems. *Nat. Electron.* **5**, 635–642 (2022).

11. Neef, V. *et al.* Three-dimensional non-Abelian quantum holonomy. *Nat. Phys.* **19**, 30–34 (2023).

12. Chen, Z.-G., Zhang, R.-Y., Chan, C. T. & Ma, G. Classical non-Abelian braiding of acoustic modes. *Nat. Phys.* **18**, 179–184 (2022).

13. Zhang, X.-L. *et al.* Non-Abelian braiding on photonic chips. *Nat. Photon.* **16**, 390–395 (2022).

14. Wilczek, F. & Zee, A. Appearance of Gauge Structure in Simple Dynamical Systems. *Phys. Rev. Lett.* **52**, 2111–2114 (1984).

15. Stern, A. Anyons and the quantum Hall effect—A pedagogical review. *Annals of Physics* **323**, 204–249 (2008).

16. Mawson, T., Petersen, T. C., Slingerland, J. K. & Simula, T. P. Braiding and Fusion of Non-Abelian Vortex Anyons. *Phys. Rev. Lett.* **123**, 140404 (2019).

17. Wu, Y., Liu, J. & Xie, X. Recent progress on non-Abelian anyons: from Majorana zero modes to topological Dirac fermionic modes. *Sci. China Phys. Mech. Astron.* **66**, 267004 (2023).

18. Klitzing, K. v., Dorda, G. & Pepper, M. New Method for High-Accuracy Determination of the Fine-Structure Constant Based on Quantized Hall Resistance. *Physical Review Letters* **45**, 494–497 (1980).

19. Haldane, F. D. M. Model for a Quantum Hall Effect without Landau Levels: Condensed-Matter Realization of the 'Parity Anomaly'. *Physical Review Letters* **61**, 2015–2018 (1988).

20. Fu, L. Topological Crystalline Insulators. *Phys. Rev. Lett.* **106**, 106802 (2011).

21. Chiu, C.-K., Teo, J. C. Y., Schnyder, A. P. & Ryu, S. Classification of topological quantum matter with symmetries. *Rev. Mod. Phys.* **88**, 035005 (2016).

22. Wu, Q., Soluyanov, A. A. & Bzdušek, T. Non-Abelian band topology in noninteracting metals. *Science* **365**, 1273–1277 (2019).

23. Guo, Q. *et al.* Experimental observation of non-Abelian topological charges and edge states. *Nature* **594**, 195–200 (2021).

24. Jiang, T. *et al.* Four-band non-Abelian topological insulator and its experimental realization. *Nat Commun* **12**, 6471 (2021).

25. Jiang, B. *et al.* Experimental observation of non-Abelian topological acoustic semimetals and their phase transitions. *Nat. Phys.* **17**, 1239–1246 (2021).

26. Wang, M. *et al.* Experimental Observation of Non-Abelian Earring Nodal Links in Phononic Crystals. *Phys. Rev. Lett.* **128**, 246601 (2022).

27. Hu, Y. *et al.* Observation of two-dimensional time-reversal broken non-Abelian topological states. *Nature Communications* **15**, 10036 (2024).

28. Sun, X.-C., Wang, J.-B., He, C. & Chen, Y.-F. Non-Abelian Topological Phases and Their Quotient Relations in Acoustic Systems. *Phys. Rev. Lett.* **132**, 216602 (2024).

29. Bouhon, A. *et al.* Non-Abelian reciprocal braiding of Weyl points and its manifestation in ZrTe. *Nat. Phys.* **16**, 1137–1143 (2020).

30. Lenggenhager, P. M., Liu, X., Tsirkin, S. S., Neupert, T. & Bzdušek, T. From triple-point materials to multiband nodal links. *Phys. Rev. B* **103**, L121101 (2021).

31. Chen, S., Bouhon, A., Slager, R.-J. & Monserrat, B. Non-Abelian braiding of Weyl nodes via symmetry-constrained phase transitions. *Phys. Rev. B* **105**, L081117 (2022).

32. Slager, R.-J., Bouhon, A. & Ünal, F. N. Non-Abelian Floquet braiding and anomalous Dirac string phase in periodically driven systems. *Nat Commun* **15**, 1144 (2024).

33. Hajdini, I. & Stoytchev, O. The Fundamental Group of $SO(n)$ Via Quotients of Braid Groups. Preprint at https://doi.org/10.48550/arXiv.1607.05876 (2016).

34. Jimbo, M. Introduction to the Yang-Baxter equation. *International Journal of Modern Physics A* **4**, 3759–3777 (1989).

35. Hatcher, A. *Algebraic Topology*. (Cambridge University Press, 2002).

36. Zak, J. Symmetry criterion for surface states in solids. *Physical Review B* **32**, 2218–2226 (1985).

37. Sun, X.-Q., Zhang, S.-C. & Bzdušek, T. Conversion Rules for Weyl Points and Nodal Lines in Topological Media. *Phys. Rev. Lett.* **121**, 106402 (2018).

38. Nelson, A., Neupert, T., Bzdušek, T. & Alexandradinata, A. Multicellularity of Delicate Topological Insulators. *Phys. Rev. Lett.* **126**, 216404 (2021).

39. Nelson, A., Neupert, T., Alexandradinata, A. & Bzdušek, T. Delicate topology protected by rotation symmetry: Crystalline Hopf insulators and beyond. *Phys. Rev. B* **106**, 075124 (2022).

40. Oka, T. & Kitamura, S. Floquet Engineering of Quantum Materials. *Annu. Rev. Condens. Matter Phys.* **10**, 387–408 (2019).

41. Li, T. & Hu, H. Floquet non-Abelian topological insulator and multifold bulk-edge correspondence. *Nature Communications* **14**, 6418 (2023).

42. Sjöqvist, E. *et al.* Non-adiabatic holonomic quantum computation. *New J. Phys.* **14**, 103035 (2012).

43. Abdumalikov Jr, A. A. *et al.* Experimental realization of non-Abelian non-adiabatic geometric gates. *Nature* **496**, 482–485 (2013).

44. Chen, Z.-G., Wang, L., Zhang, G. & Ma, G. Chiral Symmetry Breaking of Tight-Binding Models in Coupled Acoustic-Cavity Systems. *Phys. Rev. Applied* **14**, 024023 (2020).

45. Deng, Y. *et al.* Observation of Degenerate Zero-Energy Topological States at Disclinations in an Acoustic Lattice. *Phys. Rev. Lett.* **128**, 174301 (2022).

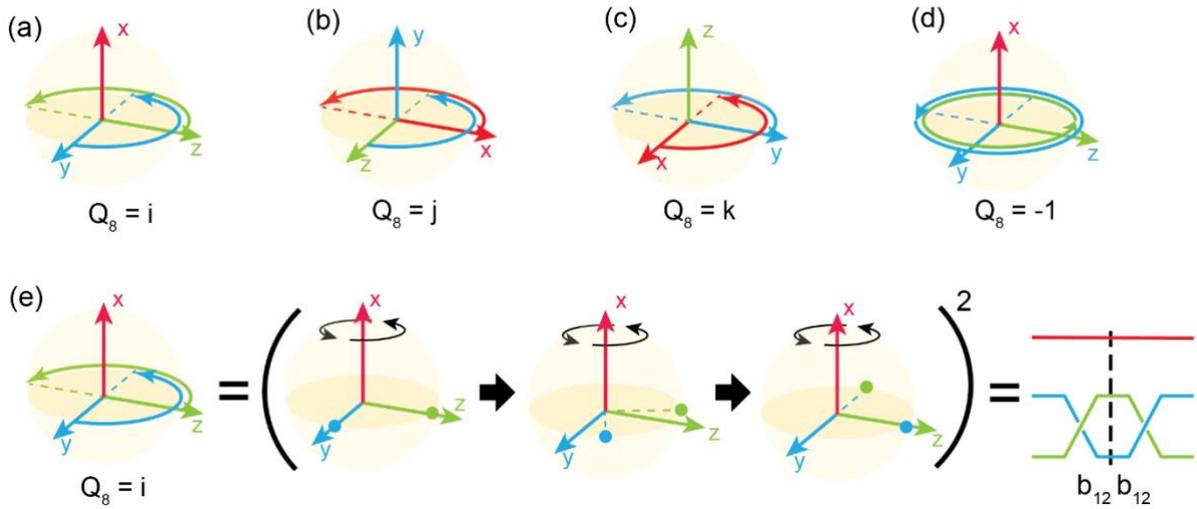

**Figure 1** (a)-(d) The eigenvector frame rotation of non-Abelian topological phases with quaternion topological invariants. (e) The case $Q_8 = i$ is characterized by the π rotation about the *x*-axis. It can be decomposed into two sequential $\pi/2$ rotations, each maps to the braiding operation of $b_{12}$. (f) The similar can be seen for $Q_8 = k$.

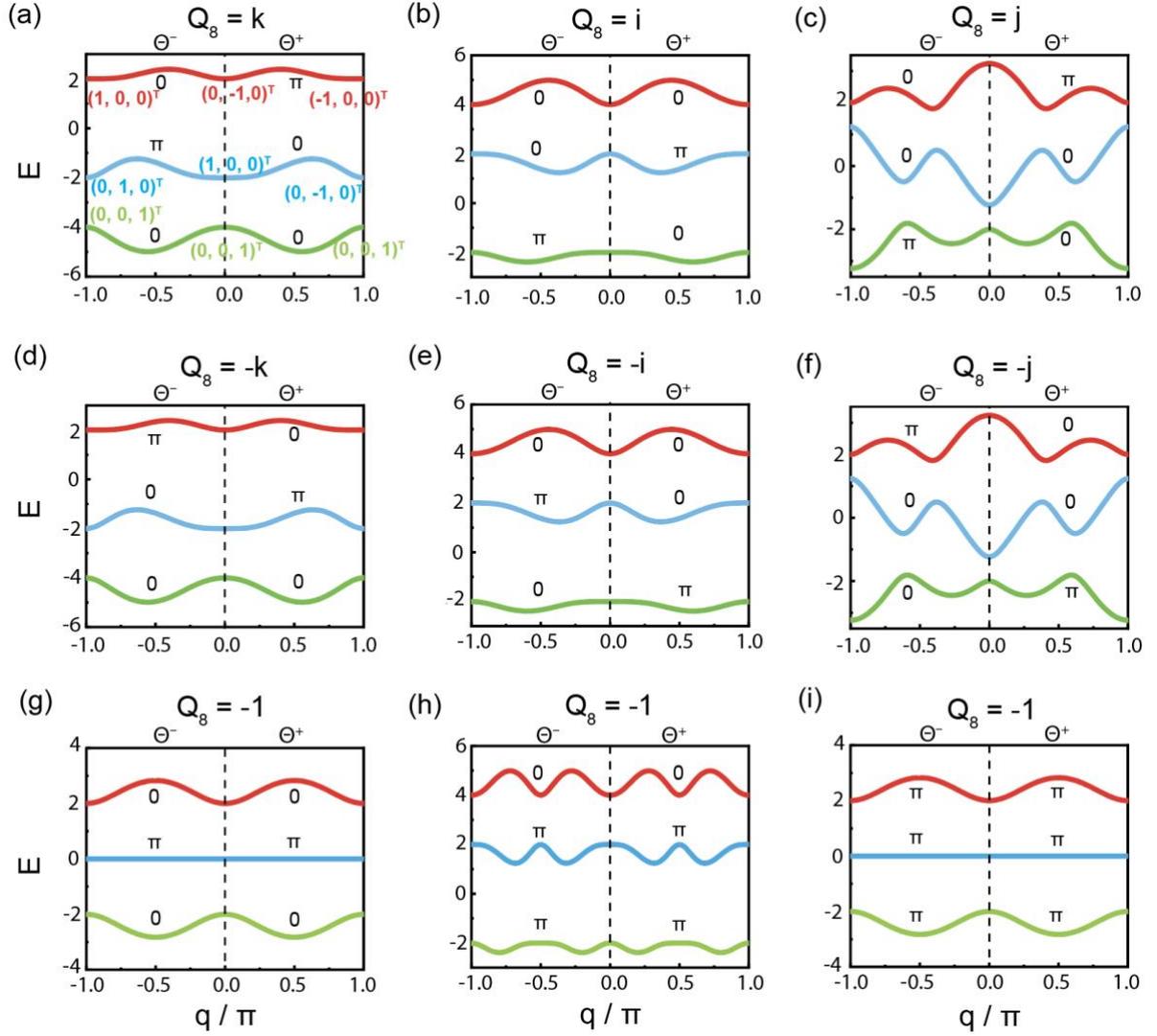

**Figure 2** The band structures, quaternion invariants, and half-BZ geometric phases of the three-band model [Eq. (6)] under different non-Abelian topological phases. In (a), the Bloch eigenvectors at the edges and center of the BZ are indicated, in which the profiles of the first (green) and second (blue) eigenvectors are braided twice, corresponding to the braiding sequence of $b_{12}b_{12}$.

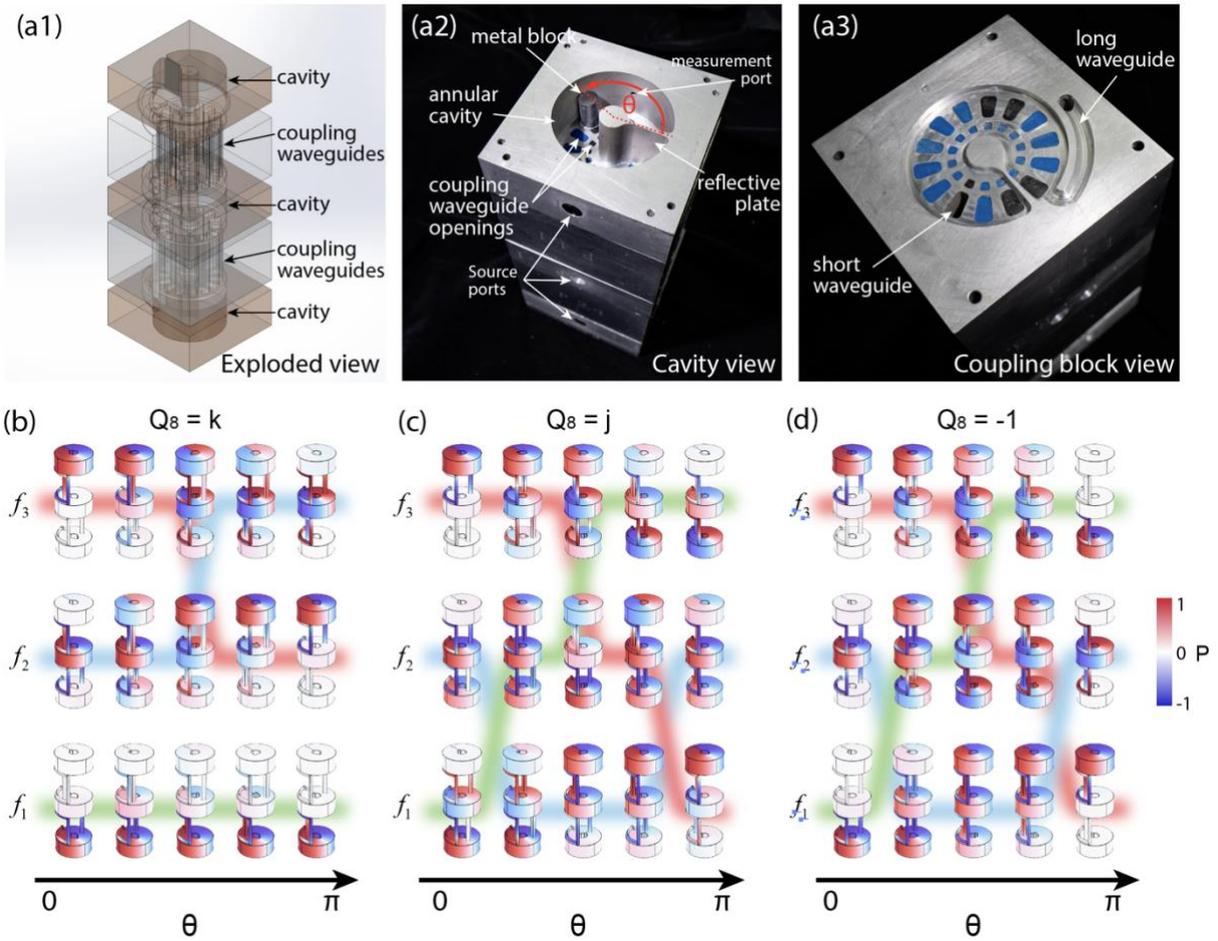

**Figure 3** (a1-a3) The design of the coupled acoustic cavities. (b-d) Numerically simulated pressure fields in the coupled acoustic cavities realizing the non-Abelian topological insulators with quaternion invariants $Q_8 = k, j, -1$, respectively. The evolutions of acoustic pressure fields (color maps) driven by $\theta$ follows the corresponding braiding rules. The colored strands in the backgrounds of (b-d) are guides to the eye.

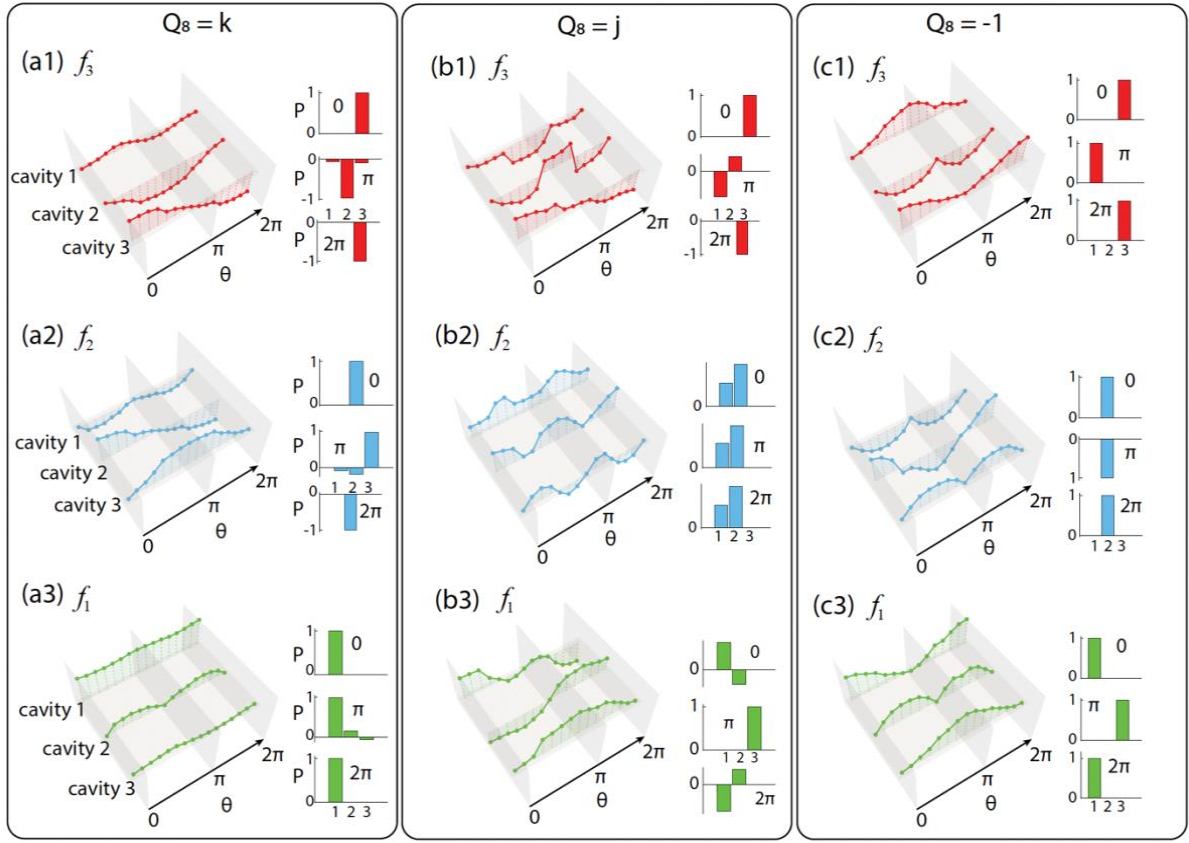

**Figure 4** Measured eigenfunctions of the coupled acoustic cavities realizing non-Abelian quaternion invariant $Q_8 = k$ (a1-a3), $Q_8 = j$ (b1-b3), and $Q_8 = -1$ (c1-c3). Each panel shows the evolution of one Bloch eigenmode from $\theta = 0$ to $2\pi$. The horizontal plane marks zero pressure. The left insets in each panel show the measured pressure distributions in the three cavities at $\theta = 0, \pi, 2\pi$.

# Supplementary Information

# Topological Braiding of Bloch Eigenmodes Protected by

# Non-Abelian Quaternion Invariants

Xiaoming Wang[1,2], Jiaying Xu[1], Xulong Wang[1], Zhen Li[1], Guancong Ma[1,3†]

[1]Department of Physics, Hong Kong Baptist University, Kowloon Tong, Hong Kong, China

[2]Department of Physics and Astronomy, The University of Tennessee, Knoxville, Tennessee 37996, USA

[3]Shenzhen Institute for Research and Continuing Education, Hong Kong Baptist University, Shenzhen 518000, China

1. **Homotopy classification of topological spaces**

In this section, we first review the foundational concepts of homotopy group and then apply this framework to classify the multiband one-dimensional (1D) *PT*-symmetric topological phase. An abbreviated derivation can also be found in the Supplementary Materials of ref. [1].

1.1. **Definition of homotopy group**

On an $n$-sphere $S^n$, we choose a based point $a$. For a topological space $M$ with a base point $b$, the $n$-th homotopy group $\pi_n(M)$ is defined as the set of equivalence classes of continuous maps $f$

$$f: S^n \to M | f(a) = b, \tag{S1}$$

which map $a$ on the sphere to the $b$ on $M$. This is equivalent to embedding the $n$-sphere into the topological space $M$ to resolve its topological properties. Two maps belong to the same homotopy class if they can continuously deform into one another with the base point fixed. To define the group operation of homotopy group, we firstly define the product $f * g$ of two loops, where $f, g$ denote two mapping with $f, g: [0,1] \to M$

$$f * g = \begin{cases} f(2t) & t \in \left[0, \frac{1}{2}\right] \\ g(2t-1) & t \in \left[\frac{1}{2}, 1\right] \end{cases}. \tag{S2}$$

If the loop $f$ and loop $g$ can continuously transform to each other, it means that $f$ and $g$ are categorized into the same homotopy class. Each distinct homotopy class denote a generators of homotopy group.

1.2. **Group Cohomology and projective representation**

In certain cases, directly computing homotopy groups using exact sequences proves challenging. To circumvent this difficulty, we extend the group structure of the topological space—a strategy that simplifies homotopy group calculations. Central to this approach is the concept of projective representations, which generalize standard linear representations of groups and are indispensable in studying symmetry-protected topological (SPT) phases. Projective representation is a generalization of the standard (linear) representation of a group. It arises in quantum mechanics and condensed matter physics, especially in the context of symmetry- protected topological (SPT) phases, where it plays a crucial role in characterizing edge states and symmetry properties. A standard representation of a group $G$ is a homomorphism $\rho: G \to GL(V)$, where where $GL(V)$ is the group of invertible linear transformations on a vector space $V$. This means

$$\rho(g_1 g_2) = \rho(g_1)\rho(g_2), \forall g_1, g_2 \in G. \tag{S3}$$

A projective representation is a map $\tilde{\rho}: G \to GL(V)$ that satisfies

$$\tilde{\rho}(g_1 g_2) = \omega(g_1, g_2)\tilde{\rho}(g_1)\tilde{\rho}(g_2), \forall g_1, g_2 \in G, \tag{S4}$$

where $\omega(g_1, g_2)$ is a scalar factor called the 2-cocycle and $\omega(g_1, g_2) \in \mathcal{H}^2(G, M)$, where $\mathcal{H}^2(G, M)$ denotes second order group Cohomoloy. Generally, $G$ is a group and $M$ is a module of $G$. It studies how $G$ act on $M$ and lead to various of structure. $M$ is an Abelian group. The group cohomology study how the group $G$ extend over module $M$. The topological phase can be differentiated by projective representation of $G$ over module $M$. Moreover, how the group $G$ is projectively represented is deterimined by its corresponding group cohomology.

### 1.3. Homotopy group of the Su-Schrieffer-Heeger model

Now we proceed to derive the homotopy group $\pi_1(X_{\text{SSH}})$. The Hamiltonian is a map from $k \in [-\pi, \pi]$ to the topological space $X_{\text{SSH}}$. Obviously, $X_{\text{SSH}} \cong S^1$. So $\pi_1(X_{\text{SSH}})$ has one generator denoted as $g$, which is the group element of $\pi_1(S^1)$. All the group elements can be directly given as $\pi_1(S^1) = \{g^n\}$, where $n = 0, \pm 1, \pm 2, \dots, \pm N$. Directly, we have $\pi_1(S^1) = \{g^n\} \cong \mathbb{Z}$, which is the topological classification of SSH model.

### 1.4. Homotopy group of 1D *PT*-symmetric systems:

The system of our interest has an $N$-band *PT*-symmetric Hamiltonian

$$H(k) = R(k)\mathcal{E}R(k)^{\text{T}}, \tag{S5}$$

where $\mathcal{E} = \text{diag}(1, 2, 3, \dots, N)$, and $R(k) \in SO(N)$. Equation S5 is fully gapped for all $k$. In the presence of *PT* symmetry, the topological space is $M = \frac{O(n)}{(\mathbb{Z}_2)^N}$, where $(\mathbb{Z}_2)^N = \mathbb{Z}_2 \times \mathbb{Z}_2 \times \dots \times \mathbb{Z}_2$ ($N$ times of cross product of $\mathbb{Z}_2$. The underlying parameter manifold of $k$ is one-dimensional (1D), so we can classify the topological space $M$ by $\pi_1(M)$. Consider the three-band case, $M = \frac{O(3)}{\mathbb{Z}_2 \times \mathbb{Z}_2 \times \mathbb{Z}_2}$, i.e., all three eigenvectors are embedded on the $O(3)$ manifold and each eigenvector has a $\mathbb{Z}_2$ gauge degree of freedom. The quotient operation is applied to eliminate additional gauge degree of freedom of $\mathbb{Z}_2$.

Now we start to derive the homotopy group $\pi_1\left(\frac{O(3)}{\mathbb{Z}_2 \times \mathbb{Z}_2 \times \mathbb{Z}_2}\right)$. Because $\det(O(3)) = \pm 1$, $O(3)$ is two disconnected manifolds with identical topology but different orientations. The redundancy can be removed by

$$\frac{O(3)}{\mathbb{Z}_2 \times \mathbb{Z}_2 \times \mathbb{Z}_2} \cong \frac{SO(3)}{\mathbb{Z}_2 \times \mathbb{Z}_2}. \tag{S6}$$

We employ the exact sequence of homotopy groups. Given a group $G$ and subgroup $F \in G$, the coset is $\frac{G}{F}$ and the exact sequence is

$$\cdots \to \pi_n(G) \to \pi_n\left(\frac{G}{F}\right) \to \pi_{n-1}(F) \to \pi_{n-1}(G) \to \cdots \tag{S7}$$

If $\pi_n(G) = \pi_{n-1}(G)$, $\pi_n\left(\frac{G}{F}\right) \cong \pi_{n-1}(F)$. This conclusion will be used latter.

For $\pi_1\left(\frac{SO(3)}{\mathbb{Z}_2 \times \mathbb{Z}_2}\right)$, we have the following sequence

$$\cdots \to \pi_1(SO(3)) \to \pi_1\left(\frac{SO(3)}{\mathbb{Z}_2 \times \mathbb{Z}_2}\right) \to \pi_0(\mathbb{Z}_2 \times \mathbb{Z}_2) \to \pi_0(SO(3)) \to \cdots \tag{S8}$$

Because $\pi_1(SO(3)) = \mathbb{Z}_2$, Eq. (S8) simplifies to

$$\cdots \to \mathbb{Z}_2 \to \pi_1\left(\frac{SO(3)}{\mathbb{Z}_2 \times \mathbb{Z}_2}\right) \to \mathbb{Z}_2 \times \mathbb{Z}_2 \to 0 \to \cdots \tag{S9}$$

So we cannot reach the conclusion that $\pi_1\left(\frac{SO(3)}{\mathbb{Z}_2 \times \mathbb{Z}_2}\right) = \mathbb{Z}_2 \times \mathbb{Z}_2$. This indicates that the element in $\pi_1\left(\frac{SO(3)}{\mathbb{Z}_2 \times \mathbb{Z}_2}\right)$ and $\mathbb{Z}_2 \times \mathbb{Z}_2$ are not one-to-one mapping. To overcome this obstacle, we perform group extension on both $SO(3)$ and $\mathbb{Z}_2 \times \mathbb{Z}_2$. and over $\mathbb{Z}_2$. The group extension is determined by following group cohomology as

$$\mathcal{H}^2(\mathbb{Z}_2 \times \mathbb{Z}_2, \mathbb{Z}_2) = \mathbb{Z}_2, \tag{S10}$$
$$\mathcal{H}^2(SO(3), \mathbb{Z}_2) = \mathbb{Z}_2 \tag{S11}$$

$\mathbb{Z}_2 \times \mathbb{Z}_2$ is isomorphic to the order-2 dihedral group $D_2 = \{1, C_{2x}, C_{2y}, C_{2z}\}$, where $C_{2x}, C_{2y}, C_{2z}$ represent two-fold rotation along $x, y, z$ axis, respectively. A nontrivial extension of $\mathbb{Z}_2 \times \mathbb{Z}_2$ is given as $D_2 \to Q_8 = \{\pm\sigma_0, \pm i\sigma_x, \pm i\sigma_y, \pm i\sigma_z\}$. The extension modifies the group relations as $C_{2x}^2 = I$ and $(\pm i\sigma_x)^2 = \pm\sigma_0$. This is the direct result of projective representation. Similarly, the nontrivial extension of $SO(3)$ is expressed as $SO(3) \to SU(2)$. Finally, we reach

$$\frac{SO(3)}{\mathbb{Z}_2 \times \mathbb{Z}_2} \cong \frac{SU(2)}{Q_8}. \tag{S12}$$

The exact sequence of $\pi_1\left(\frac{SU(2)}{Q_8}\right)$ is

$$\cdots \to \pi_1(SU(2)) \to \pi_1\left(\frac{SU(2)}{Q_8}\right) \to \pi_0(Q_8) \to \pi_0(SU(2)) \to \cdots \tag{S13}$$

Note that $\pi_1(SU(2)) = \pi_0(SU(2)) = 0$, such that we can reach $\pi_1\left(\frac{SU(2)}{Q_8}\right) \cong \pi_0(Q_8) = Q_8$ In summary, the band topology of a three-band $PT$-symmetric Hamiltonian is classified by the quaternion group.

Extending to four-band case, the topological space is $X = \frac{SO(4)}{(\mathbb{Z}_2)^4}$. Repeat the same procedure, we firstly extend the group $SO(4)$ and $(\mathbb{Z}_2)^4$ over the module $\mathbb{Z}_2$. Then we obtain the following exact sequence

$$\cdots \to \pi_1(SU(3)) \to \pi_1\left(\frac{SU(3)}{Q_{16}}\right) \to \pi_0(Q_{16}) \to \pi_0(SU(3)) \to \cdots \tag{S14}$$

where $Q_{16}$ is the generalized quaternion group (or the octonion group). Because $\pi_1(SU(3)) = \pi_0(SU(3)) = 0$, $\pi_1\left(\frac{SU(3)}{Q_{16}}\right) \to \pi_0(Q_{16}) = Q_{16}$.

## 2. Relative homotopy group and the half-BZ geometric phase

Topological phases of matter are traditionally determined by *homotopy groups*, which gives topological invariants defined over full BZ, such as Zak phase, Chern number, or $\mathbb{Z}_2$ topological invariant. When a system has symmetry-invariant sub-topological spaces, such as boundaries or symmetric planes, the *relative homotopy groups* is a useful tool to reveal topological information that can be missed by homotopy groups. It has recently been successfully applied to analyze delicate topological insulator[2]. The half-BZ geometric phase associated with non-Abelian band topology in our work is also defined and determined using the relative homotopy groups.

The relative homotopy group, denoted $\pi_n(X, A)$, studies how maps from an $n$-dim disk $D^n$ (or an $n$-ball) into a space $X$ behave relative to a subspace $A \subset X$. Formally, we define a continuous map $f: (D^n, S^{n-1}) \to (X, A)$, i.e., $f$ maps the interior of $D^n$ is mapped into $X$, and the boundary of of $D^n$, which is an $(n-1)$-dim sphere $S^{n-1}$, to $A \subset X$. We then consider the homotopy class of such maps, where the deformation keeps the boundary inside $A$.

The difference between homotopy group and relative homotopy group are the followings.
- An $n$-th homotopy group: $f: S^n \to M$ maps an $n$-sphere to a topological space, where $S^n$ is compact and has no boundary.
- In contrast, an $n$-th relative homotopy group: $f: (D^n, S^{n-1}) \to (X, A)$ embed a disk $D^n$, which has an $S^{n-1}$ boundary, into the topological space
- The relations between the homotopy group and relative homotopy group is linked by a long exact sequence

$$\cdots \xrightarrow{\partial_{n+1}} \pi_n(A) \xrightarrow{i_n} \pi_n(X) \xrightarrow{j_n} \pi_n(X, A) \xrightarrow{\partial_n} \pi_{n-1}(A) \xrightarrow{\partial_{n-1}} \cdots \tag{S15}$$

An example that applies the relative homotopy group to classify this topological phase is reported in ref.[2], we reproduce it here for the readers' convenience. Considering a three-dimensional Hamiltonian $h(\mathbf{k})$ with mirror symmetry $R_z h(\mathbf{k}) R_z^{-1} = h(R_z \mathbf{k})$, where

$$R_z(k_x, k_y, k_z) \to (k_x, k_y, -k_z). \tag{S16}$$

In other words, $h(\mathbf{k})$ is invariant on the $k_x k_y$-plane at $k_z = 0, \pi$, i.e.,

$$R_z h(k_x, k_y 0) R_z^{-1} = h(k_x, k_y, 0). \tag{S17}$$

The topological space of $h(\mathbf{k})$ and $h(k_x, k_y, 0)$ are denoted $X$ and $A$, respectively. Obviously, $A \subset X$. The relative homotopy group have a nontrivial topological classification $\pi_2(X, A) = \mathbb{Z}$. This indicate that there is a nontrivial map

$$D^2 \to X \tag{S18}$$

Directly, we can define a Chern number over the upper and lower half BZ

$$C^+ = \int_{D^2} \Omega^+(\boldsymbol{k}) d\boldsymbol{k}, \tag{S19}$$

$$C^- = \int_{D^2} \Omega^-(\boldsymbol{k}) d\boldsymbol{k}, \tag{S20}$$

where $\Omega^\pm(\boldsymbol{k})$ represent the Berry curve on the upper and lower half BZ. $C^+$ and $C^-$ defined over the two halves of the BZ form a Berry dipole. It is straightforward to check that the net Chern number of the full BZ is zero, so the topological structure of the Berry dipole is missed by the homotopy group.

Now we return to the system in the main text. Our models [Eq. (1), or more concretely, Eq. (6)] also have a mirror-symmetry-invariant topological subspace. We apply the relative homotopy group to study the band topology of the quaternion phases. The topological space of $H(q)$ is given as

$$X = \frac{O(3)}{O(1) \times O(1) \times O(1)}. \tag{S21}$$

The mirror-symmetric points are $q = 0$ and $\pi$. The topological space of $H(q)$ at $q = 0$ and $\pi$ are the boundary subspace $A$. As reported in ref.[1], the relative homotopy group is

$$\pi_1(X, A) = \mathbb{Z}_2. \tag{S22}$$

This nontrivial result permits the definition of a half-BZ geometric phase,

$$\Theta_n^- = \text{Im} \log\langle\psi_n(-\pi)|\psi_m(0)\rangle, \quad \Theta_n^+ = \text{Im} \log\langle\psi_m(0)|\psi_n(\pi)\rangle. \tag{S23}$$

In stark contrast with the single-band geometric (Zak) phase defined over full BZ, $\Theta_n^\pm$ cannot be determined by considering a single band, as indicated by the different band indices $m$ and $n$. The indices are chosen according to the braid sequence to satisfy the holonomy condition over half BZ. In other words, $|\psi_n(-\pi)\rangle$ and $|\psi_m(0)\rangle$ (also $|\psi_m(0)\rangle$ and $|\psi_n(\pi)\rangle$) must have an identical eigenfunction (up to a phase factor). And the result given by Eq. (S22) ensures that $\Theta_n^\pm$ quantize to 0 or $\pi$ over half BZ.

### 3. Homotopy loop on $SO(3)$ manifold

In this section, we establish connections between the representations of Artin's braiding group and homotopy loops on the $SO(N)$ manifold. The discussion is primarily informed by the research in ref.[3] The first homotopy group of $SO(N)$ is $\pi_1(SO(N)) = \mathbb{Z}/2\mathbb{Z} = \mathbb{Z}_2$ ($N \geq 3$). To enumerate all the possible homotopy loops on the $SO(3)$ sphere, we build a mapping as

$$f: [0,1] \to SO(3), \tag{S24}$$

which embed the $S^1$ on $SO(3)$. This mapping the same as Eq. (S2), which corresponds to a homotopy loop, where the base point and end point are given by $f(0)$ and $f(1)$, respectively. The positions of the base and end points can be represented by three-dimensional vectors, denoted as $V(0)$ and $V(1)$. Ref.[3] has shown that all possible homotopy loops on the $SO(N)$ manifold can be expressed as a product of a series of braiding operations, leading to the relation

$$V(1) = \tau V(0), \tag{S25}$$

where $\tau$ denotes a braiding sequence. To enumerate all possible homotopy loops on $SO(3)$ within the framework of braiding operations, ref.[3] proposes using a set of generators $g$ to construct the

braiding sequence

$$\tau = g_i g_k \cdots g_l, \tag{S26}$$

where the subscripts $i, j, k$, etc., label different generators here. The generators effectively map continuous rotations into discrete braiding operations. These generators form a group $G$, given by

$$G = \{b_{12}^m b_{23}^n, b_{12}^m b_{23} b_{12}, b_{12}^m b_{23}^3 b_{12}\}, \tag{S27}$$

where $m = 0,1,\ldots,7, n = 0,1,2,3$. The group elements satisfy the below equivalence relations

$$b_{12} b_{23} b_{12} = b_{23} b_{12} b_{23}, \quad b_{12}^2 = b_{23} b_{12}^2 b_{23}, \quad b_{12}^8 = b_{23}^8 = 1. \tag{S28}$$

The group $G$ consists of 48 elements. This is to say, any homotopy on the SO(3) sphere can adiabatically evolve into one of these homotopy loops described above.

Extending this framework to higher-dimensional spaces, i.e., SO($N$), the number of generators in the group $G$ is given by $2^N N!$. In Section 1.3, we derived the following sequence:

$$\cdots \to \pi_1(\mathrm{SO}(3)) \to \pi_1\left(\frac{\mathrm{SO}(3)}{\mathbb{Z}_2 \times \mathbb{Z}_2}\right) \to \pi_0(\mathbb{Z}_2 \times \mathbb{Z}_2) \to \pi_0(\mathrm{SO}(3)) \to \cdots \tag{S29}$$

with $\pi_1\left(\frac{\mathrm{SO}(3)}{\mathbb{Z}_2 \times \mathbb{Z}_2}\right) = Q_8$. Thus, every $\mathbb{Z}_2$ homotopy loop can be assigned to one of the $Q_8$ homotopy loops.

In the presence of mirror symmetry, we impose the condition $\exp\left[\int_{-\pi}^{0} A_{3\times 3}(q) dq\right] = \exp\left[\int_{0}^{\pi} A_{3\times 3}(q) dq\right]$. This constraint ensures that $\exp\left[\int_{-\pi}^{0} A_{3\times 3}(q) dq\right] \in G$. Consequently, the homotopy loops must respect mirror symmetry. The sequences satisfying this requirement are

$$\tau = g^2. \tag{S30}$$

These generators form a subset of $G$, denoted as $G_m$, which is explicitly given by

$$G_m = \{b_{12}^{\pm 1}, b_{23}^{\pm 1}, b_{12} b_{23} b_{12}^{-1}, b_{12} b_{23}^{-1} b_{12}^{-1}, b_{12}^{\pm 2}, b_{23}^{\pm 2}, b_{12} b_{23} b_{12}, b_{12} b_{23}^{-1} b_{12},$$

$$b_{12}^{\pm 1} b_{23}^{\pm 2}, b_{23}^{\pm 1} b_{12}^{\pm 2}, b_{12} b_{23}^{-1} b_{12}, b_{12} b_{23}^{-1} b_{12}\}. \tag{S31}$$

Let us denote an element of $G_m$ as

$$g_m = \exp\left[\int_{-\pi}^{0} A_{3\times 3}(\theta) d\theta\right]. \tag{S32}$$

Notably, the elements of $G_m$ do not form a group. The braiding sequence $g_m^2$ maps to one of the eight quaternion charges. All possible irreducible braiding sequences are listed in Table S1. Additionally, elements in the set $G_m$ can be transformed into group elements in $G$ through equivalent transformations, even if they initially appear different.

Table S1. Relation between the quaternion charges, the NABPs, and braid sequences.

| $Q_8$ | Non-Abelian Berry phase $W$ | Braid word |
|---|---|---|
| $i$ | $\exp(\pi L_x) = \mathrm{diag}(1,-1,-1)$ | $b_{12}^2$ |
| $-i$ | $\exp(\pi L_z) = \mathrm{diag}(1,-1,-1)$ | $b_{12}^{-2}$ |

| $k$ | $\exp(\pi L_z) = \text{diag}(-1, -1, 1)$ | $b_{23}^2$ |
|---|---|---|
| $-k$ | $\exp(-\pi L_z) = \text{diag}(-1, -1, 1)$ | $b_{23}^{-2}$ |
| $j$ | $\exp(\pi L_y) = \text{diag}(-1, 1, -1)$ | $b_{12}b_{23}b_{12}^{-1}b_{12}b_{23}b_{12}^{-1}$ |
| $-j$ | $\exp(-\pi L_y) = \text{diag}(-1, 1, -1)$ | $b_{12}b_{23}^{-1}b_{12}^{-1}b_{12}b_{23}^{-1}b_{12}^{-1}$ |
| $-1$ | $\text{diag}(1, 1, 1)$ | $b_{12}^4$ |
| | | $b_{23}^4$ |
| | | $(b_{12}b_{23}b_{12}^{-1})^4$ |
| | | $(b_{12}b_{23}^{-1}b_{12}^{-1})^4$ |
| | | $(b_{12}b_{23}^2)^4$ |
| | | $(b_{23}b_{12}^2)^4$ |
| | | $(b_{12}b_{23}b_{12})^2$ |
| | | $(b_{12}b_{23}^{-1}b_{12})^2$ |
| | | $(b_{23}b_{12}b_{23})^2$ |
| | | $(b_{23}b_{12}^{-1}b_{23})^2$ |

## 4. Generic way to realize arbitrary braiding sequence.

The non-Abelian topological models can be generalized to produce arbitrary braiding sequences. The generalization consists of two steps. The first step is to show that similar braiding phenomena exist beyond three-band models. For example, a four-band *PT*-symmetric Hamiltonian is categorized by a generalized quaternion group (also called the octonion group)

$$Q_{16} := \left\{ \begin{matrix} \pm 1, \pm q_{12}, \pm q_{23}, \pm q_{34}, \pm q_{13}, \pm q_{24}, q_{1234}, -q_{1234} \mid \\ (\pm q_{ij})^2 = -1, \ q_{ij}q_{kl} = -q_{kl}q_{ij}, q_{13} = q_{12}q_{23}, q_{24} = q_{23}q_{34}, q_{1234} = q_{12}q_{23}q_{34} \end{matrix} \right\}. \quad (S33)$$

The complete multiplication table are provided in ref.[4].

The evolution of such a four-band model follows $SO(4)$, which has six generators as

$$J_{12} = \begin{pmatrix} 0 & 0 & 0 & 0 \\ 0 & 0 & 0 & 0 \\ 0 & 0 & 0 & -1 \\ 0 & 0 & 1 & 0 \end{pmatrix}, J_{23} = \begin{pmatrix} 0 & 0 & 0 & 0 \\ 0 & 0 & -1 & 0 \\ 0 & 1 & 0 & 0 \\ 0 & 0 & 0 & 0 \end{pmatrix}, J_{34} = \begin{pmatrix} 0 & -1 & 0 & 0 \\ 1 & 0 & 0 & 0 \\ 0 & 0 & 0 & 0 \\ 0 & 0 & 0 & 0 \end{pmatrix},$$

$$J_{13} = \begin{pmatrix} 0 & 0 & 0 & 0 \\ 0 & 0 & 0 & -1 \\ 0 & 0 & 0 & 0 \\ 0 & 1 & 0 & 0 \end{pmatrix}, J_{24} = \begin{pmatrix} 0 & 0 & -1 & 0 \\ 0 & 0 & 0 & 0 \\ 1 & 0 & 0 & 0 \\ 0 & 0 & 0 & 0 \end{pmatrix}, J_{14} = \begin{pmatrix} 0 & 0 & 0 & -1 \\ 0 & 0 & 0 & 0 \\ 0 & 0 & 0 & 0 \\ 1 & 0 & 0 & 0 \end{pmatrix}. \quad (S34)$$

Similar to lifting the $SO(3)$ to $SU(2)$, a similar group lifting consists of the following mappings

$$J_{12} \to -\tfrac{i}{2}\lambda_3, J_{23} \to -\tfrac{i}{2}\lambda_2, J_{34} \to -\tfrac{i}{2}\lambda_1, J_{13} \to -\tfrac{i}{2}\lambda_6, J_{24} \to -\tfrac{i}{2}\lambda_5, J_{14} \to -\tfrac{i}{2}\lambda_4, \quad (S35)$$

where $\lambda_i$ are the generators of $SU(3)$, i.e., the Gell-Mann matrices,

$$\lambda_1 = \begin{pmatrix} 0 & 1 & 0 \\ 1 & 0 & 0 \\ 0 & 0 & 0 \end{pmatrix}, \lambda_2 = \begin{pmatrix} 0 & -i & 0 \\ i & 0 & 0 \\ 0 & 0 & 0 \end{pmatrix}, \lambda_3 = \begin{pmatrix} 1 & 0 & 0 \\ 0 & -1 & 0 \\ 0 & 0 & 0 \end{pmatrix},$$

$$\lambda_4 = \begin{pmatrix} 0 & 0 & 1 \\ 0 & 0 & 0 \\ 1 & 0 & 0 \end{pmatrix}, \lambda_5 = \begin{pmatrix} 0 & 0 & i \\ 0 & 0 & 0 \\ -i & 0 & 0 \end{pmatrix}, \lambda_6 = \begin{pmatrix} 0 & 0 & 0 \\ 0 & 0 & 1 \\ 0 & 1 & 0 \end{pmatrix},$$

$$\lambda_7 = \begin{pmatrix} 0 & 0 & 0 \\ 0 & 0 & -i \\ 0 & i & 0 \end{pmatrix}, \lambda_8 = \frac{1}{\sqrt{3}} \begin{pmatrix} 1 & 0 & 0 \\ 0 & 1 & 0 \\ 0 & 0 & -2 \end{pmatrix}. \tag{S36}$$

After lifting from SO(4) to SU(3), the NABP of generalized quaternion charge can be quantized to be $\pm i\lambda_{1-6}, \pm \lambda_3 \lambda_1$. We build a mapping from NABP to $Q_{16}$ as $\pm i\lambda_3 \to \pm q_{12}, \pm i\lambda_2 \to \pm q_{23}, \pm i\lambda_1 \to \pm q_{34}, \pm i\lambda_6 \to \pm q_{13}, \pm i\lambda_5 \to \pm q_{24}, \pm i\lambda_4 \to \pm q_{14}, \pm \lambda_3 \lambda_1 \to \mp q_{1234}$. As stated in ref.[4], the homotopy loop on SO(4) sphere can be also decomposed as the multiplication of braiding operation. The corresponding group $G$ has 384 ($2^4 \times 4!$) elements[4]. However, some of the braiding sequences do not have the mirror symmetry. Therein, the braid operators protected by mirror symmetry have the below representations:

$$b_{12} \to \begin{pmatrix} 1 & 0 & 0 & 0 \\ 0 & 1 & 0 & 0 \\ 0 & 0 & 0 & -1 \\ 0 & 0 & 1 & 0 \end{pmatrix}, b_{23} \to \begin{pmatrix} 1 & 0 & 0 & 0 \\ 0 & 0 & -1 & 0 \\ 0 & 1 & 0 & 0 \\ 0 & 0 & 0 & 1 \end{pmatrix}, b_{34} \to \begin{pmatrix} 0 & -1 & 0 & 0 \\ 1 & 0 & 0 & 0 \\ 0 & 0 & 1 & 0 \\ 0 & 0 & 0 & 1 \end{pmatrix}. \tag{S37}$$

**Table S2.** Relation between $Q_{16}$, non-Abelian Berry phase, and braid sequences in $B_4$. We only list part of the braiding sequences. Particularly, generalized quaternion invariant of $-1$ have corresponds to many different braiding sequences. We do not enumerate all cases for the brevity of this document.

| $Q_{16}$ | Non-Abelian Berry phase | Braid word |
| --- | --- | --- |
| $-1$ | $\exp(\pm 2\pi J_{ij})$ | $b_{12}^{\pm 4}, b_{23}^{\pm 4}, b_{34}^{\pm 4}$ |
| $\pm q_{12}$ | $\exp(\pm \pi J_{12})$ | $b_{12}^{\pm 2}, (b_{12} b_{23}^{\pm 2})^{\mp 2}$ |
| $\pm q_{23}$ | $\exp(\pm \pi J_{23})$ | $b_{23}^{\pm 2}$ |
| $\pm q_{34}$ | $\exp(\pm \pi J_{34})$ | $b_{34}^{\pm 2},, (b_{23}^{\pm 2} b_{34})^{\mp 2}$ |
| $\pm q_{13}$ | $\exp(\pm \pi J_{13})$ | $(b_{12} b_{23} b_{12}^{-1})^{\pm 2}$ |
| $\pm q_{24}$ | $\exp(\pm \pi J_{24})$ | $(b_{23} b_{34} b_{23}^{-1})^{\pm 2}$ |
| $\pm q_{14}$ | $\exp(\pm \pi J_{14})$ | $(b_{12} b_{23} b_{34} b_{23}^{-1} b_{12}^{-1})^{\pm 2}$ |
| $q_{1234}$ | $\exp(\pi J_{12} + \pi J_{34})$ | $b_{12}^2 b_{34}^2$ |
| $-q_{1234}$ | $\exp(\pi J_{12} - \pi J_{34})$ | $b_{12}^2 b_{34}^{-2}$ |

In the second step, we present the method of producing arbitrary braid sequences. In the main text, all braid sequences are produced by solely relying on the properties of the Bloch modes and Bloch bands. And the possibility of braid sequences, although plentiful, are still limited. This hurdle can be overcome by employing piecewise evolution of the $n$-band Hamiltonian

$$H(t) = R(t)\mathcal{E}R^{\mathrm{T}}(t) \tag{S38}$$

where $\mathcal{E}$ is a diagonal matrix with $\mathcal{E} = \mathrm{diag}(1,2,3,\ldots,N)$ and $t$ is time (or some other driving parameter). As long as $R(t) \in \mathrm{SO}(N)$, the system remains fully gapped and the evolution is protected by non-Abelian band topology. We use four-band model, i.e., $\mathcal{E} = \mathrm{diag}(1,2,3,4)$ as an example to illustrate how to realize $b_{12}b_{23}b_{34}$. This braid sequence has an odd number of crossings and cannot be produced by the closed-loop evolution. The full evolution is composed of three segments,

$$R(t) = \begin{cases} \exp(tJ_{12}) & t \in \left[0, \frac{\pi}{2}\right] \\ \exp\left[\left(t - \frac{\pi}{2}\right)J_{23}\right] & t \in \left(\frac{\pi}{2}, \pi\right] \\ \exp\left[\left(t - \frac{3\pi}{2}\right)J_{34}\right] & t \in \left(\pi, \frac{3\pi}{2}\right] \end{cases} \tag{S39}$$

where $J_{ij}$ represent the generators of $\mathrm{SO}(4)$. The NABPs at critical points $t = \frac{\pi}{2}, \pi, \frac{3\pi}{2}$ are derived by $W_{t'} = \mathcal{P} \exp\left[\int_0^{t'} A_{4\times 4}(t)dt\right]$, and they are

$$W_{\frac{\pi}{2}} = \begin{pmatrix} 1 & 0 & 0 & 0 \\ 0 & 1 & 0 & 0 \\ 0 & 0 & 0 & -1 \\ 0 & 0 & 1 & 0 \end{pmatrix} \to b_{12},$$

$$W_{\pi} = \begin{pmatrix} 1 & 0 & 0 & 0 \\ 0 & 0 & -1 & 0 \\ 0 & 0 & 0 & -1 \\ 0 & 1 & 0 & 0 \end{pmatrix} \to b_{12}b_{23},$$

$$W_{\frac{3\pi}{2}} = \begin{pmatrix} 0 & -1 & 0 & 0 \\ 0 & 0 & -1 & 0 \\ 0 & 0 & 0 & -1 \\ 1 & 0 & 0 & 0 \end{pmatrix} \to b_{12}b_{23}b_{34}. \tag{S40}$$

As shown in Ref. [3], it has been proven that all homotopy classes on $\mathrm{SO}(N)$ sphere can be expressed as a product of a series of braiding sequences. This lays the foundation for extending the approach to higher dimensions.

5. **Robustness of non-Abelian braiding**

The braiding of Bloch modes is protected by non-Abelian band topology, therefore it is robust against random perturbations. We use the $j$-phase as an example to demonstrate the robustness. Perturbations are introduced as $H(q) + \delta H$, where $H(q)$ is the Bloch Hamiltonian [Eq. (6) in the main text] and $\delta H$ is a $3 \times 3$ random matrix with elements being random numbers in $[-V, V]$. The parameters are set as $\omega_1 = \omega_2 = \omega_3 = 0, v_3 = 1, v_2 = 0, v_1 = -1, w_1 = 0, w_2 = u = 1$. The evolution of the eigenvectors is plotted in Fig. S1 for different $V$. It is clearly seen that although the perturbation introduces fluctuations in the trajectory, the overall sequence remains intact. Note that if the system is clean, the point that splits the non-Abelian Berry phase equally is at the middle of the Brillouin zone (BZ). In the presence of disorder that breaks the mirror symmetry, such a point drifts from the center. But since the evolution is a holonomy, such a point always exists even in the presence of disorder.

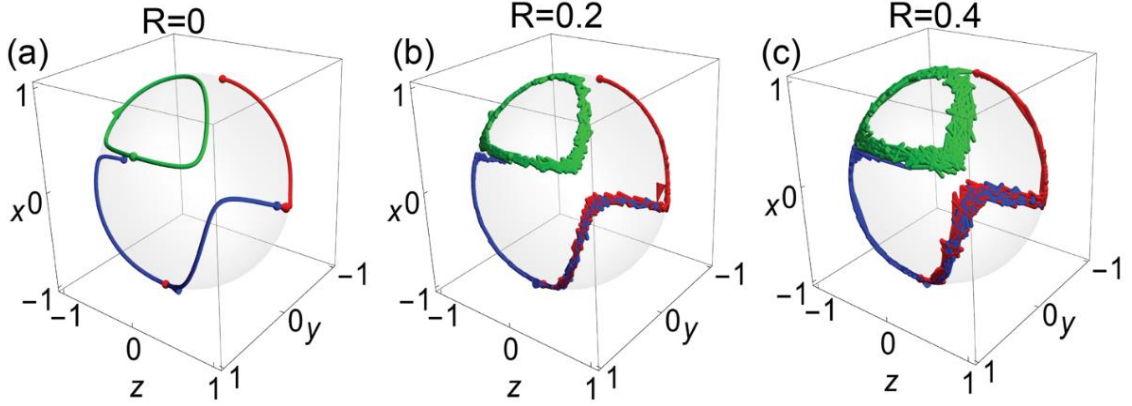

**Figure S1.** Eigenvector evolution of non-Abelian quaternion charge $Q_8 = j$ in the presence of disorder. The red, green, and blue lines represent the evolution of the eigenvectors corresponding to the third, second, and first eigenmodes, respectively.

## 6. The coupled acoustic cavities

We use a two-level model to illustrate the design of the acoustic cavities and the realization of the synthetic dimension $\theta$. The cavities have an annular cross-section and are coupled by two small waveguides of different lengths. A thin metal plate is inserted in each cavity to reflect the circulating traveling waves, such that azimuthal standing-wave modes are formed. The first resonance is used as the onsite orbital. Its pressure field profile is shown in Fig. S2(b).

In Eq. (6), the onsite energies are controlled by the Bloch wavenumber $q$, which maps to the azimuthal coordinate $\theta$. This is realized by varying the azimuthal position of a small metal block in the cavity, which tunes the resonant frequency, as shown in Fig. S2(b).

Observe that in Eq. (6), when driving the Bloch $q$ across the BZ, the coupling coefficients may need to smoothly go through zero. According to previous studies, the sign of the coupling coefficient is tunable by varying the length of coupling waveguide[5]. In our design, two neighboring cavities are coupled with two waveguides with different lengths. The short (straight) one provides negative coupling [Type-I in Fig. S2(a)], and the long one (bended, thrice the length of the short waveguide) provides positive coupling [Type-II in Fig. S2(a)]. In the experiments, the azimuthal position of the short waveguide is varied and the long waveguides is fixed. The change of coupling coefficients manifests as the change in the splitting of the two coupled modes, as shown in Fig. S2(c, d). In Fig. S2(c), only the short waveguide exists, and the two coupled modes do not flip in parity, so there is no sign change in the coupling coefficient. In Fig. S2(d), the two modes smoothly cross twice, indicating the change coupling sign.

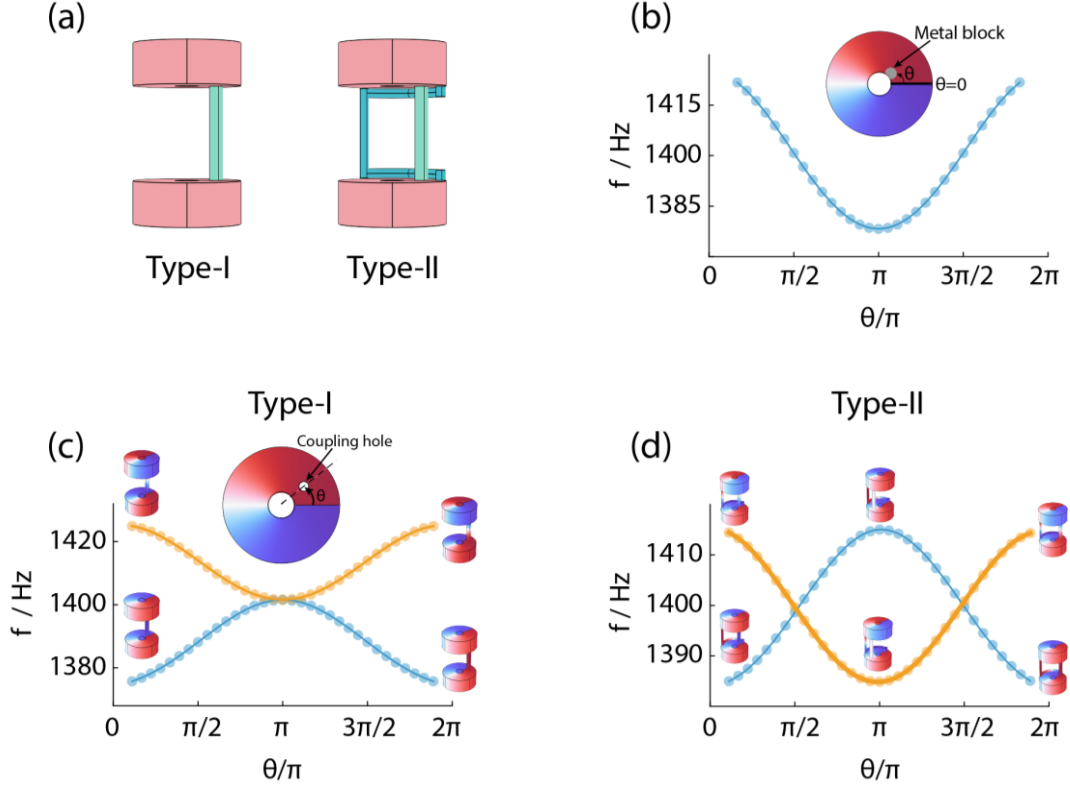

**Figure S2.** (a) Acoustic cavities coupled by two waveguides. Type-II is the experimental configuration. (b) The resonant frequency of a single cavity as a function of the azimuthal position of the metal block. (c) Resonant frequencies of two cavities connected in Type-I configuration. (d) Resonant frequencies of two cavities connected in Type-II configuration. Insets show the pressure field profiles of resonant modes, with red/white/blue representing positive/zero/negative pressure.

### 7. Parameters used in numerical calculation in the main text

Table S3 lists all the parameters of the tight-binding model. The numerical results in Figure 2 of the main text can be reproduced using the parameters provided in Table S3. The non-Abelian quaternion charge -1 allows for multiple braiding sequences, which are detailed in the last three rows of Table S3. The labels $-1(g)$, $-1(h)$, and $-1(i)$ correspond to the parameters required to generate the results shown in Fig. 2 (g), (h), (i) of the main text, respectively. The parameters used in finite-element simulations (COMSOL Multiphysics) are listed in Table S4.

**Table S3.** Parameters in tight-biding calculations (Fig. 2 in the main text).

| $Q_8$ | $\omega_1$ | $\omega_2$ | $\omega_3$ | $v_1$ | $v_2$ | $v_3$ | $w_1$ | $w_2$ | $u$ |
|---|---|---|---|---|---|---|---|---|---|
| $i$ | 0 | 0 | 4 | $-1$ | 1 | 0 | 1 | 0 | $-1$ |
| $-i$ | 0 | 0 | 4 | $-1$ | 1 | 0 | $-1$ | 0 | 1 |
| $k$ | $-4$ | 0 | 0 | 0 | $-1$ | 1 | $-1$ | 0 | 1 |
| $-k$ | $-4$ | 0 | 0 | 0 | $-1$ | 1 | $-1$ | 0 | 1 |
| $j$ | 0 | 0 | 0 | $-1$ | 0 | 1 | 0 | 1 | 1 |
| $-j$ | 0 | 0 | 0 | $-1$ | 0 | 1 | 0 | 1 | $-1$ |
| $-1$(g) | 0 | 0 | 0 | $-1$ | 0 | 1 | 1 | 0 | $-1$ |
| $-1$(h) | 0 | 0 | 4 | $-1$ | 1 | 0 | 1 | 0 | $-1$ |
| $-1$(i) | 0 | 0 | 0 | $-1$ | 0 | 1 | 1 | 0 | 1 |

Table S4.  Parameters used in numerical simulations ($rad \cdot s^{-1}$).

| $Q_8$ | $\omega_1$ | $\omega_2$ | $\omega_3$ | $v_1$ | $v_2$ | $v_3$ | $w_1$ | $w_2$ | $u$ |
|---|---|---|---|---|---|---|---|---|---|
| $k$ | 8610 | 8792.2 | 8792.2 | 0 | $-94.2$ | 94.2 | $-80.4$ | 0 | 80.4 |
| $j$ | 8792.2 | 8792.2 | 8792.2 | $-94.2$ | 0 | 94.2 | 0 | 80.4 | 80.4 |
| $-1$ | 8792.2 | 8792.2 | 8792.2 | $-94.2$ | 0 | 94.2 | $-80.4$ | 0 | 80.4 |

## 7. Experimental parameters

The Hamiltonian in Eq.(6) can be reformulated as

$$H = \begin{pmatrix} \omega_3 - i\Gamma_0 & t_1 & 0 \\ t_1 & \omega_2 - i\Gamma_0 & t_2 \\ 0 & t_2 & \omega_1 - i\Gamma_0 \end{pmatrix} \quad (S41)$$

The three coupled cavities, interconnected via acoustic tubes, are accurately modeled by the simplified Hamiltonian presented above. In some special configurations, we impose the condition $t_1 = -t_2$ to reduce the degree of freedoms. The parameters (e.g., coupling strengths $t_1$, $t_2$) and eigenvectors are determined by fitting experimental data using the Green's function method, which has been described in the main text. The fitting parameters are listed in Tables S5-S7. Figure S3 displays some selective measured responses.

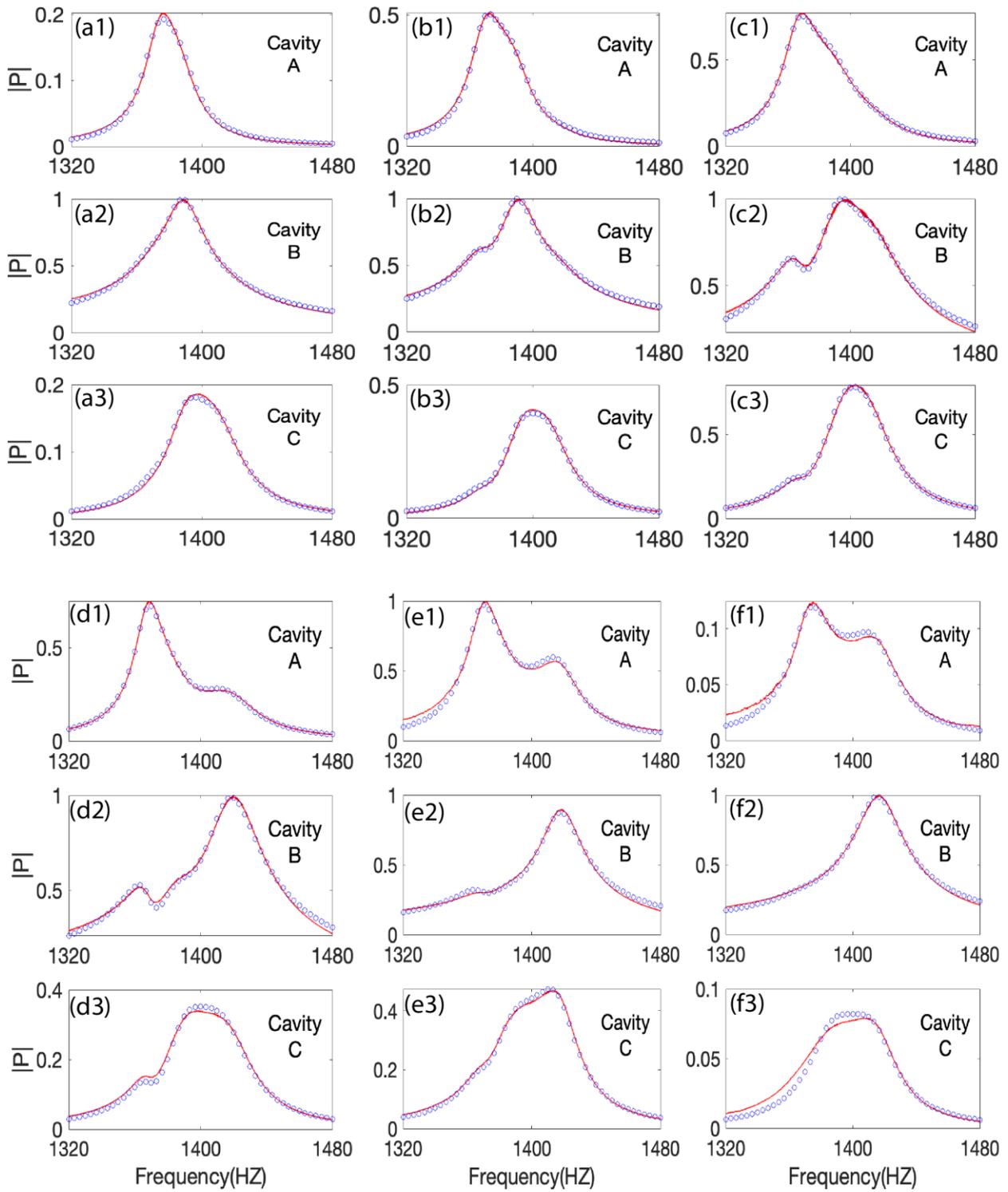

Figure S3. Measured and fitted frequency response of three coupled acoustic cavities. The microphone was at the second cavity. The markers show the experimental measurement, and the red solid cruves represent the Green's function fitting results. Parameters for panels (a)–(f) are listed in Table S5 corresponding to Points No. 2–7 respectively.

**Table S5.** The fitting parameters of the acoustic system for $Q_8 = k$. Here $t_2 = -t_1$, and units of all parameters are $\mathrm{rad \cdot s^{-1}}$.

| Points No. | $\omega_1$ | $\omega_2$ | $\omega_3$ | $t_2$ |
|---|---|---|---|---|
| 1 | 8635.1 | 8701.9 | 8890.4 | 0 |
| 2 | 8627.6 | 8715.1 | 8876.3 | 32.7 |
| 3 | 8615.2 | 8729.1 | 8862.6 | 49.4 |
| 4 | 8609.1 | 8761.4 | 8838.1 | 72.6 |
| 5 | 8617.1 | 8839.6 | 8756.6 | 79.6 |
| 6 | 8614.9 | 8860.2 | 8724.8 | 71.0 |
| 7 | 8619.5 | 8881.4 | 8702.8 | 34.6 |
| 8 | 8635.1 | 8890.4 | 8701.9 | 0 |
| 9 | 8634.2 | 8901.2 | 8723.5 | −39.9 |
| 10 | 8605.2 | 8864.8 | 8741.9 | −63.7 |
| 11 | 8616.7 | 8839.9 | 8765.5 | −69.1 |
| 12 | 8610.5 | 8769.7 | 8829.5 | −75.4 |
| 13 | 8606.4 | 8741 | 8859.5 | −69.1 |
| 14 | 8617.4 | 8706.2 | 8880.3 | −39.9 |
| 15 | 8617.6 | 8684.3 | 8901.2 | −18.8 |
| 16 | 8635.1 | 8701.9 | 8890.4 | 0 |

**Table S6.** The fitting parameters of the acoustic system for $Q_8 = j$. Units of all parameters are $\mathrm{rad \cdot s^{-1}}$.

| Points No. | $\omega_1$(rad/s) | $\omega_2$(rad/s) | $\omega_3$(rad/s) | $t_1$(rad/s) | $t_2$(rad/s) |
|---|---|---|---|---|---|
| 1 | 8690.1 | 8792.5 | 8887.6 | 81.1 | 0 |
| 2 | 8716.1 | 8803.0 | 8884.6 | 76.0 | 26.2 |
| 3 | 8734.3 | 8806.8 | 8865.8 | 59.7 | 53.7 |
| 4 | 8759.8 | 8807.6 | 8774.1 | 27.5 | 75.5 |
| 5 | 8817.6 | 8804.2 | 8780.3 | −13.9 | 79.1 |
| 6 | 8866.2 | 8804.9 | 8730.8 | −59.7 | 53.8 |
| 7 | 8885.9 | 8805.2 | 8714.9 | −76.0 | 26.2 |
| 8 | 8890.1 | 8797.4 | 8689.1 | −80.5 | 0 |
| 9 | 8878.3 | 8810.1 | 8720.8 | −69.6 | −40.2 |
| 10 | 8837.4 | 8802.1 | 8763.7 | −33.9 | −72.8 |
| 11 | 8798.9 | 8804.7 | 8780.5 | −1.4 | −80.3 |
| 12 | 8771.7 | 8805.6 | 8825.4 | 26.1 | −76.0 |
| 13 | 8756.0 | 8807.1 | 8847.3 | 40.2 | −69.6 |
| 14 | 8722.6 | 8800.2 | 8880.1 | 70.9 | −37.7 |

| 15 | 8690.1 | 8792.5 | 8887.6 | 81.1 | 0 |

**Table S7.** The fitting parameters of the acoustic system for $Q_8 = -1$. Here $t_1 = t_2$, and units of all parameters are $\text{rad} \cdot \text{s}^{-1}$.

| Points No. | $\omega_1$(rad/s) | $\omega_2$(rad/s) | $\omega_3$(rad/s) | $t_1$(rad/s) |
|---|---|---|---|---|
| 1 | 8706.5 | 8800.0 | 8894.8 | 0 |
| 2 | 8714.3 | 8799.1 | 8884.4 | 34.2 |
| 3 | 8743.3 | 8804.0 | 8857.9 | 63.3 |
| 4 | 8769.4 | 8798.7 | 8829.8 | 76.1 |
| 5 | 8829.2 | 8799.5 | 8769.4 | 77.4 |
| 6 | 8857.4 | 8796.7 | 8743.9 | 64.9 |
| 7 | 8880.9 | 8804.1 | 8718.3 | 38.6 |
| 8 | 8894.0 | 8800.0 | 8705.5 | 0 |
| 9 | 8880.4 | 8803.2 | 8717.9 | −38.9 |
| 10 | 8847.8 | 8801.4 | 8754.8 | −69.5 |
| 11 | 8803.9 | 8898.3 | 8800.1 | −81.2 |
| 12 | 8774.3 | 8797.9 | 8821.8 | −77.7 |
| 13 | 8736.9 | 8800.2 | 8861.3 | −62.9 |
| 14 | 8714.6 | 8799.6 | 8889.4 | −28.5 |
| 15 | 8706.5 | 8800.0 | 8894.8 | 0 |

**References**

1. Wu, Q., Soluyanov, A. A. & Bzdušek, T. Non-Abelian band topology in noninteracting metals. *Science* **365**, 1273–1277 (2019).
2. Sun, X.-Q., Zhang, S.-C. & Bzdušek, T. Conversion Rules for Weyl Points and Nodal Lines in Topological Media. *Phys. Rev. Lett.* **121**, 106402 (2018).
3. Hajdini, I. & Stoytchev, O. The Fundamental Group of $SO(n)$ Via Quotients of Braid Groups. Preprint at https://doi.org/10.48550/arXiv.1607.05876 (2016).
4. Jiang, T. *et al.* Four-band non-Abelian topological insulator and its experimental realization. *Nat Commun* **12**, 6471 (2021).
5. Chen, Z.-G., Wang, L., Zhang, G. & Ma, G. Chiral Symmetry Breaking of Tight-Binding Models in Coupled Acoustic-Cavity Systems. *Phys. Rev. Applied* **14**, 024023 (2020).